\documentclass[journal,comsoc]{IEEEtran}

\usepackage[T1]{fontenc}% optional T1 font encoding

\usepackage{times}

\usepackage{amsmath}  % Define \boldsymbol (in amsbsy too) and align
\usepackage{amssymb}  % Define \mathbb (in amsfonts too)
\usepackage{mathrsfs} % Define \mathscr, a script font

\usepackage{theorem}  % Helps in rendering theorems, etc.
\usepackage{cite}     % Gives multiple s as intervals
\usepackage{comment}  % Comments out with \begin{comment} ... \end{comment}
\usepackage{multirow}

\usepackage{upref}
\usepackage{amsfonts}

\usepackage{verbatim}

\usepackage[dvipsnames,usenames]{color}

\usepackage{graphicx}
\usepackage{subfigure}
\usepackage{tikz}
\usetikzlibrary{shapes,snakes}
\usetikzlibrary{arrows,positioning} 
\usepackage{psfrag}
\usepackage{float}
\restylefloat{table}
\usepackage{algorithmicx}
\usepackage{algorithm}
\usepackage{algpascal}
\usepackage{algcompatible}
\usepackage{algpseudocode}
\usepackage{accents}

\usepackage{epsfig}
\newcommand{\bc}{\begin{center}}
\newcommand{\ec}{\end{center}}

\newcommand{\cC}{{\cal C}}

\newcommand{\cN}{{\cal N}}

\newcommand{\bfe}{{\boldsymbol e}}

\newcommand{\bfp}{{\boldsymbol p}}

\newcommand{\bfr}{{\boldsymbol r}}
\newcommand{\bfs}{{\boldsymbol s}}

\newcommand{\bfx}{{\boldsymbol x}}

\newcommand{\bfz}{{\boldsymbol z}}

\newcommand{\bfJ}{{\mathbf J}}

\renewcommand{\leq}{\leqslant}

\renewcommand{\geq}{\geqslant}

\newcommand{\Cref}[1]{Co\-rol\-la\-ry\,\ref{#1}}

\newcommand{\Figref}[1]{Fig.~\ref{#1}}

\newcommand{\Equref}[1]{equation (\ref{#1})}

\theoremstyle{plain} \theorembodyfont{\normalfont\slshape}

\newtheorem{thm}{Theorem$\!$}

\newtheorem{prop}{Proposition$\!$}
\newenvironment{proposition}{\begin{prop}\hspace*{-1ex}{\bf.}}{ \end{prop}}

\newtheorem{lem}{Lemma$\!$}

\newtheorem{conj}[thm]{Conjecture$\!$}

\newtheorem{cor}{Corollary$\!$}

\newtheorem{cl}{Claim$\!$}

\newtheorem{defi}{Definition$\!$}

\newtheorem{const}{Construction$\!$}

\theorembodyfont{\normalfont}

\newtheorem{exam}{Example$\!$}
\newenvironment{example}{\begin{exam}\hspace*{-1ex}{\bf .}}{\end{exam}}

\newtheorem{remrk}{Remark$\!$}
\newenvironment{remark}{\begin{remrk}\hspace*{-1ex}{\bf .}}{\end{remrk}}

\newlength{\paragraphindent}
\newlength{\widthone}
\newlength{\widthtwo}
\newlength{\widththree}
\newlength{\colwidthtemp}
\definecolor{Codecolor}{named}{White}  %{Tan}
\newcommand{\Copen}{\mbox{\{\kern-5.50pt\{}}
\newcommand{\Cclose}{\mbox{\}\kern-5.50pt\}}}
\newcommand{\Cslash}{\mbox{$\backslash\kern-6.02pt\backslash$}}

\begin{document}
\title{Multihead Multitrack Detection with Reduced-State Sequence Estimation}
\author{Bing Fan,~\IEEEmembership{Student Member,~IEEE,}
        Hemant K. Thapar,~\IEEEmembership{Fellow,~IEEE,}
        and Paul H. Siegel,~\IEEEmembership{Fellow,~IEEE}% <-this % stops a space
\thanks{B. Fan and P. H. Siegel are with the Department of Electrical \& Computer
Engineering, University of California, San Diego, CA 92093 USA (e-mail:
bifan@ucsd.edu; psiegel@ucsd.edu)}% <-this % stops a space
\thanks{H. Thapar is with OmniTier Storage (e-mail: hemantkthapar@gmail.com)}}% <-this % stops a space
%\thanks{Manuscript received ..., 2016; revised ..., 2016.}}

%\markboth{IEEE Transactions on Communications,~Vol., No., Jan~2016}%
%{}

\maketitle
%\IEEEtitleabstractindextext{
\begin{abstract}
To achieve ultra-high storage capacity, the data tracks are squeezed more and more on the magnetic recording disks, causing severe intertrack interference (ITI). The multihead multitrack (MHMT) detector is proposed to better combat ITI. Such a detector, however, has prohibitive implementation complexity. In this paper we propose to use the reduced-state sequence estimation (RSSE) algorithm to significantly reduce the complexity, and render MHMT practical. We first consider a commonly used symmetric two-head two-track (2H2T) channel model. The effective distance between two input symbols is redefined. It provides a better distance measure and naturally leads to an unbalanced set partition tree. Different trellis configurations are obtained based on the desired performance/complexity tradeoff. Simulation results show that the reduced MHMT detector can achieve near maximum-likelihood (ML) performance with a small fraction of the original number of trellis states. Error event analysis is given to explain the behavior of RSSE algorithm on 2H2T channel. Search results of dominant RSSE error events for different channel targets are presented. We also study an asymmetric 2H2T system. The simulation results and error event analysis show that RSSE is applicable to the asymmetric channel.

\end{abstract}
\begin{IEEEkeywords}
Shingled Magnetic Recording, Multitrack Multihead Detection, Intertrack Interference, Reduced-State Sequence Estimation
\end{IEEEkeywords}
%}

\IEEEpeerreviewmaketitle

% % % % % % % % % % % % %Tikz practice % % % % % % % % % % % % % % %

% % % % % % % % % % % % % % % % SECTION 1  % % % % % % % % % % % % % % % % % % % % % % %
\section{Introduction}
\label{sec_intro}

Intertrack interference (ITI), caused by aggressively shrinking the track pitch, is one of the more severe impairments in next generation hard disk drives (HDDs) \cite{Wood_TDMR}\cite{KSW2008}. The use of an array reader to simultaneously read and process multiple tracks has recently drawn intensive interest because of its capability to handle ITI as well as electronic noise \cite{MHPGH2014}\cite{Xia2015}. The associated maximum likelihood (ML) detector complexity is, however, drastically increased. 

Consider a symmetric two-head two-track (2H2T) system described by
\begin{align}
\left[\! \begin{array}{c}
r^a(D) \\ r^b(D)
\end{array} \!\right]
= \left[ \!\begin{array}{c c}
1 & \epsilon \\ \epsilon & 1
\end{array}\! \right]
\left[ \!\begin{array}{c}
x^a(D)h(D) \\ x^b(D)h(D)
\end{array}\! \right] + 
\left[ \!\begin{array}{c}
n^a(D) \\ n^b(D)
\end{array} \!\right],
\label{eq_1}
\end{align}
where $x^a(D)$, $x^b(D)$ are the data sequences independently recorded on two adjacent tracks $a$ and $b$, with $x^i(D)=\Sigma_{k=0}^N\,x_k^iD^k$ and $x_k^i\in\{-1,+1\}$ for $i\in\{a,b\}$. Each single track is equalized to the same target $h(D)=h_0+h_1D+\dots+h_{\nu}D^{\nu}$, and the ITI effect is characterized by a parameter $\epsilon$. The received sequences from two heads, $r^a(D)$ and $r^b(D)$, include additive electronic noise, $n^a(D)$ and $n^b(D)$. We assume $n^{a}(D)$ and $n^{b}(D)$ are uncorrelated and i.i.d, with $n_{k}^{a},n_{k}^{b}\sim\cN(0,\sigma^{2})$. 

The corresponding ML detector simultaneously decodes two tracks by searching a joint trellis \cite{Soljanin_multihead}\cite{Ma_iterative}\cite{barbosa_simultaneously}.  Let $\bfx_k = (x_k^a, x_k^b)$ denote the $k$th input symbol of the 2H2T system. The possible input symbols form a two dimensional 4-symbol constellation. This extended input set causes exponential increase in the computation complexity. For a channel with memory $\nu$, the ML trellis has $4^{\nu}$ states, each with $4$ incoming and outgoing edges. Compared to the traditional single-head single-track (SHST) detector with complexity $O(2^v)$, the ML 2H2T detector operates at $O(4^\nu)$. For $\nu>3$, which is typical in practical recording channels, the ML 2H2T detector becomes impractical. 

In this work we consider the design of reduced-complexity detectors while retaining good performance. The underlying idea is to drop less possible paths at early stages. This low complexity algorithm, called reduced-state sequence estimation (RSSE) \cite{rsse88}, was first designed for transmitting signals from a large quadrature amplitude modulation (QAM) constellation through a partial response channel with long memory. The RSSE trellis, which was originally constructed based on the Ungerboeck set partition tree, has fewer states, but retains a well-defined structure. Efforts were made to develop similar algorithms for the 2H2T system. The authors of \cite{KPS1999} showed a way to design low complexity trellis, but their construction generally suffers from high performance loss. We propose a different approach in this paper. We find that by a simple eigen decomposition, the 2H2T system can be transformed to a QAM-type system, and the reduced-state trellis can be constructed by redefining the distance measure on the transformed input constellation. This new construction rule provides greater flexibility in performance/complexity tradeoffs. Our simulation results show that, with fewer than half the number of the states of the full ML trellis, RSSE can achieve near-ML performance on many channels. This work was partially presented in \cite{bing_intermag}.

Moreover, the evaluation of RSSE performance is tractable through error events analysis. In contrast to the ML detector, some error events in RSSE are merged early due to the reduced-state. An early-merging condition is introduced to identify these error events, and a modified error state diagram is used to search for the dominant early-merged error events. The search results on several reduced-state trellis configurations at different ITI levels are presented. When the minimum distance parameter of the early-merged error events is larger than that of the ML detector, the performance loss of RSSE trellis is almost negligible.

An asymmetric 2H2T system, where the ITI levels sensed by two heads are different, is also considered because of its practical interest. We provide a distance analysis for both the ML detector and the RSSE detector. It is shown that the proposed reduced-state trellis construction rule is also applicable in the asymmetric case.

The paper is organized as follows. In Section \ref{sec_rsse} we first briefly review the traditional RSSE algorithm for the QAM system. Next we show how to construct a reduced-state trellis for the 2H2T channel by redefining the distance measure in the input constellation and designing proper set partitioning trees. In Section \ref{sec_simulation} we construct different trellis configurations based on the performance/complexity tradeoff, and simulate the RSSE detector on several channels with different channel polynomials. The early-merging condition and error event analysis are presented in Section \ref{sec_errorevent}. The dominant error events for several reduced-state trellises on different channels are also tabulated. In Section \ref{sec_asymmetric} we consider to apply the RSSE algorithm to the asymmetric 2H2T model. We present the simulation results as well as error event analysis. The paper is concluded in Section \ref{sec_conclusion}.

\section{2H2T Detector with RSSE}
\label{sec_rsse}
\subsection{Review of RSSE}
\label{subsec_rsse}

The traditional RSSE is designed for transmitting QAM symbols through an ISI channel with channel memory $\nu$ \cite{rsse88}. Recall that in the ML detector, the trellis state is represented as a length $\nu$ vector, 
\begin{align}
\bfp_n=[\, \bfx_{n-1}, \bfx_{n-2}, \dots, \bfx_{n-\nu}\,],
\end{align}
where each symbol $\bfx_{n-k}$ is complex-valued, and selected from a two-dimensional signal set $\cC$ whose size is $M$. In RSSE, to reduce the number of trellis states, several ML states are grouped into a {\bfseries subset state}. To do this, for the $k$th element $\bfx_{n-k}$ in $\bfp_n$, a set partition $\Omega(k)$ of $\cC$ is defined, and $\bfx_{n-k}$ is represented by its subset index $a_{n-k}(k)$ in $\Omega(k)$. Notice that $\Omega(k)$ can be different for $k=1, \cdots, \nu$. Let $J_k=|\Omega(k)|$ be the number of subsets in partition $\Omega(k)$, $1\leq J_{k}\leq M$. Then the subset index $a_{n-k}(k)$ can take its value from $0,1, \cdots, J_k-1$. The corresponding subset state of $\bfp_n$ is denoted by
\begin{align}
\bfs_n=[\,a_{n-1}(1), a_{n-2}(2), \dots, a_{n-\nu}(\nu)\,].
\end{align}
The trellis constructed from all possible $\bfs_n$ is called the \textbf{subset trellis}. To obtain a well-defined trellis structure, the partition $\Omega(k)$ is restricted to be a further partition of the subsets in $\Omega(k+1)$, for $1\leq k \leq \nu-1$. This condition guarantees that for a given state $\bfs_n$ and current input $\bfx_n$, the next subset state is uniquely determined and represented as 
\begin{align}
\bfs_{n+1}=[\,a_{n}(1), a_{n-1}(2), \dots, a_{n-\nu+1}(\nu)\,],
\end{align}
where $a_{n}(1)$ is the subset index of $\bfx_{n}$ in $\Omega(1)$, $a_{n-1}(2)$ is the index of $\bfx_{n-1}$ in $\Omega(2)$, and so on.
The number of states in the subset trellis is $\prod_{k=1}^{\nu}J_k$. The complexity of a RSSE trellis can be controlled by specifying $\nu$ parameters, $J_k$ for $1\leq k \leq \nu$. We define the \textbf{configuration} of a subset trellis to be a vector $\bfJ=[J_1, J_2, \dots, J_\nu]$. A valid configuration satisfies $J_1\geq J_2\geq \dots \geq J_\nu$.

To apply the Viterbi algorithm (VA) on a subset trellis, a decision feedback scheme is introduced to calculate the branch metric, since the subset state $\bfs_n$ does not uniquely specify the most recent $\nu$ symbols. During the detection process, a modified path history is used to store the survivor symbol $\hat{\bfx}_{n-1}$ that leads to state $\bfs_n$. The actual survivor ML state $\hat{\bfp}_n$ is obtained by tracing back $\nu$ steps in the path history. We say that $\hat{\bfp}_n$ is the only one survivor ML state at time $n$ among all possible $\bfp_n$'s whose corresponding subset state is $\bfs_n$. An example will be given in the following section to illustrate this process. Error propagation may occur, but its effect is negligible \cite{rsse88}\cite{Sheen_rsse}. 

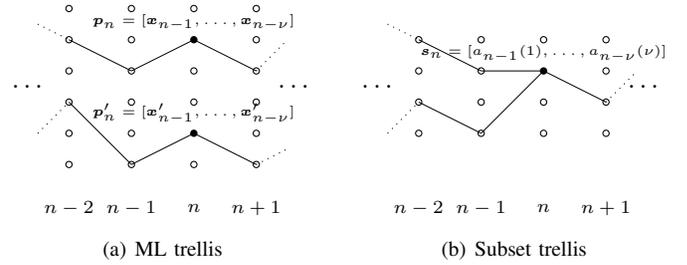
\begin{figure}
\centering
\subfigure[ML trellis]{
\begin{tikzpicture}[scale=0.83]
\draw (0,0) circle [radius=0.05];
\draw (0,0.5) circle [radius=0.05];
\draw (0,1.5) circle [radius=0.05];
\draw (0,1) circle [radius=0.05];
\draw (0,-0.5) circle [radius=0.05];
\draw (0,-1) circle [radius=0.05];
\draw (1,0) circle [radius=0.05];
\draw (1,0.5) circle [radius=0.05];
\draw (1,1.5) circle [radius=0.05];
\draw (1,1) circle [radius=0.05];
\draw (1,-0.5) circle [radius=0.05];
\draw (1,-1) circle [radius=0.05];
\draw (2,0) circle [radius=0.05];
\draw (2,0.5) circle [radius=0.05];
\draw (2,1.5) circle [radius=0.05];
\draw[fill] (2,1) circle [radius=0.05];
\draw[fill] (2,-0.5) circle [radius=0.05];
\draw (2,-1) circle [radius=0.05];
\draw (3,0) circle [radius=0.05];
\draw (3,0.5) circle [radius=0.05];
\draw (3,1) circle [radius=0.05];
\draw (3,-0.5) circle [radius=0.05];
\draw (3,-1) circle [radius=0.05];
\draw (3,1.5) circle [radius=0.05];
\node[left] at (-0.2, 0.25) { $\dots$};
\node[right] at (3.2, 0.25) { $\dots$};
\draw[dotted] (-0.5, 1.25) -- (0, 1);
\draw[dotted] (-0.5, -0.5) -- (0, 0);
\draw[very thin] (0,0) -- (1, -1);
\draw[very thin] (0,1) -- (1, 0.5);
\draw[very thin] (1,0.5) -- (2, 1);
\draw[very thin] (1,-1) -- (2, -0.5);
\draw[very thin] (2,1) -- (3, 0.5);
\draw[very thin] (2,-0.5) -- (3, -1);
\draw[dotted] (3, 0.5) -- (3.5, 1);
\draw[dotted] (3, -1) -- (3.5, -0.75);

\node[above] at (2, 1) {\tiny $\bfp_n=[\bfx_{n-1},\dots, \bfx_{n-\nu} ]$};
\node[above] at (2, -0.5) {\tiny $\bfp'_n=[\bfx'_{n-1},\dots, \bfx'_{n-\nu} ]$};

\node at (2, -1.7) {\scriptsize $n$};
\node at (3, -1.7) {\scriptsize $n+1$};
\node at (1, -1.7) {\scriptsize $n-1$};
\node at (0, -1.7) {\scriptsize $n-2$};
\end{tikzpicture}
\label{subfig_ml}
}~
\subfigure[Subset trellis]{
\begin{tikzpicture}[scale=0.83]
\draw (0,0) circle [radius=0.05];
\draw (0,0.5) circle [radius=0.05];
\draw (0,1) circle [radius=0.05];
\draw (0,-0.5) circle [radius=0.05];

\draw (1,0) circle [radius=0.05];
\draw (1,0.5) circle [radius=0.05];
\draw (1,1) circle [radius=0.05];
\draw (1,-0.5) circle [radius=0.05];

\draw (2,0) circle [radius=0.05];
\draw[fill] (2,0.5) circle [radius=0.05];
\draw (2,1) circle [radius=0.05];
\draw (2,-0.5) circle [radius=0.05];

\draw (3,0) circle [radius=0.05];
\draw (3,0.5) circle [radius=0.05];
\draw (3,1) circle [radius=0.05];
\draw (3,-0.5) circle [radius=0.05];

\node[left] at (-0.2, 0.25) { $\dots$};
\node[right] at (3.2, 0.25) { $\dots$};
\draw[dotted] (-0.5, 1.25) -- (0, 1);
\draw[dotted] (-0.5, -0.5) -- (0, 0);
\draw[very thin] (0,0) -- (1, -0.5);
\draw[very thin] (0,1) -- (1, 0.5);
\draw[very thin] (1,0.5) -- (2, 0.5);
\draw[very thin] (1,-0.5) -- (2, 0.5);
\draw[very thin] (2,0.5) -- (3, 0);
\draw[dotted] (3, 0) -- (3.5, 0.5);
\node[above] at (2, 0.5) {\tiny $\bfs_n=[a_{n-1}(1),\dots, a_{n-\nu}(\nu) ]$};
\node at (2, -1.7) {\scriptsize $n$};
\node at (3, -1.7) {\scriptsize $n+1$};
\node at (1, -1.7) {\scriptsize $n-1$};
\node at (0, -1.7) {\scriptsize $n-2$};
\end{tikzpicture}
\label{subfig_subset}
}
\caption{Comparison between the decoding paths on (a) ML trellis and (b) subset trellis. If at time $n$ two paths ending at ML states $\bfp_n$ and $\bfp_n'$ satisfy $\bfx_{n-k}\in a_{n-k}(k)$ and $\bfx'_{n-k}\in a_{n-k}(k)$ for all $k=1, \dots, \nu$, then they will merge early at subset state $\bfs_n$ in the subset trellis.
 If by time $n$ the merged paths are separated by large Euclidean distance, the early merge can be considered to be reliable since the dropped path has a high metric that makes it much less likely to be chosen as the final survivor path.}
\label{fig_rsse_ml}
\end{figure}

The underlying idea of RSSE is to drop less likely paths early in the detection process. Since each subset state contains multiple ML states, certain paths will merge earlier in the subset trellis than in the ML trellis, as shown in \Figref{fig_rsse_ml}. If $J_k=M$ for $1\leq k \leq \nu$, RSSE becomes MLSE. Otherwise it is suboptimal. To minimize the performance loss, proper set partitions $\Omega(k)$ should be selected carefully to guarantee that enough distance differences have been accumulated to reliably distinguish between merging paths. For the $M$-QAM system, it is suggested that good performance can generally be obtained by maximizing the minimum intrasubset Euclidean distance for each partition $\Omega(k)$, $k=1,\cdots, \nu$ \cite{rsse88}. The Ungerboeck set partition tree \cite{ungerboeck} is shown to have this property and is adopted to make the selection of $\Omega(k)$. For more details about the subset trellis construction for the $M$-QAM system, the reader is referred to \cite{rsse88}.

The use of the Ungerboeck set partition tree is key to obtaining good performance of the RSSE algorithm on the QAM system. However, such a set partition tree cannot be directly applied to the 2H2T system because of the ITI. In the next subsection we will show that a simple transformation can decompose the original 2H2T system into two independent channels, resulting in a QAM-like structure. Then, instead of using the Euclidean distance, we define a new distance measure between the input symbols, based on which we construct a more suitable set partition tree for the 2H2T system.

\subsection{Set Partition Tree for 2H2T System}
\label{subsec_wssjd}
In \cite{Fan1506} we show that the 2H2T channel described by \Equref{eq_1} is equivalent to
\begin{align}
\left[\! \begin{array}{c}
 r^+(D) \\ r^-(D)
\end{array} \!\right]
= 
\left[ \!\begin{array}{c}
z^+(D)h(D) \\ z^-(D)h(D)
\end{array}\! \right] + 
\left[ \!\begin{array}{c}
n^+(D) \\ n^-(D)
\end{array} \!\right],
\label{eq_minus}
\end{align}
 where 
 \begin{align}
\begin{bmatrix}
  z^+_k \\ z^-_k
\end{bmatrix}
 & = 
\begin{bmatrix}
 1 & 1\\ 1& -1
\end{bmatrix}
\begin{bmatrix}
 x^a_k \\ x^b_k
\end{bmatrix} \label{eq_z}
 \\
\begin{bmatrix}
r^+_k \\ r^-_k
\end{bmatrix}
& = 
\begin{bmatrix}
\frac{1}{1+\epsilon} & 0 \\ 0 & \frac{1}{1-\epsilon}
\end{bmatrix}
\begin{bmatrix}
1 & 1\\ 1& -1
\end{bmatrix}
\begin{bmatrix}
r^a_k \\ r^b_k
\end{bmatrix}  \label{eq_r}
\\
\begin{bmatrix}
n^+_k \\ n^-_k
\end{bmatrix}
& = 
\begin{bmatrix}
\frac{1}{1+\epsilon} & 0 \\ 0 & \frac{1}{1-\epsilon}
\end{bmatrix}
\begin{bmatrix}
1 & 1\\ 1& -1
\end{bmatrix}
\begin{bmatrix}
n^a_k \\ n^b_k
\end{bmatrix}. \label{eq_n}
 \end{align} 
%\begin{align}
% r^+(D)=\frac{r^a(D)+r^b(D)}{1+\epsilon}, \quad r^-(D)=\frac{r^a(D)-r^b(D)}{1-\epsilon}\, \notag
%\\ n^+(D)=\frac{n^a(D)+n^b(D)}{1+\epsilon}, \quad n^-(D)=\frac{n^a(D)-n^b(D)}{1-\epsilon}\,\notag
%\\ z^+(D)=x^a(D)+x^b(D), \quad z^-(D)=x^a(D)-x^b(D) \notag
% \end{align}

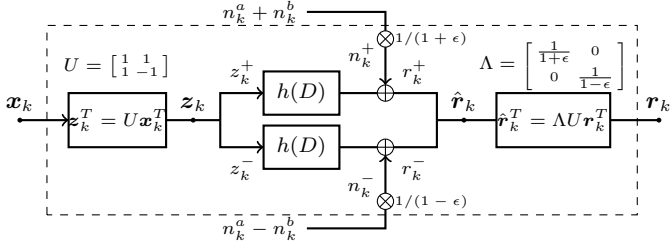
\begin{figure}
\centering
\begin{tikzpicture} [scale=0.72]
%\node at (-3.8, 0.5) {\small \color{blue} \textbf{2H2T - transformed}};
\draw [dashed] (-2.7, -1.25) rectangle (8.2, 2.25);
\draw [fill] (-3.2, 0.5) circle [radius=1pt];
\draw [thick, ->] (-3.2, 0.5) -- (-2.3, 0.5);
\draw [thick] (-2.3, 0) rectangle (-0.5, 1);
\draw [thick] (-0.5, 0.5) -- (0, 0.5);
\draw [thick] (0, 0.5) -- (0.5, 0.5);
\draw [->, thick] (0.5, 0.5) -- (0.5, 1) -- (1.3, 1);
\draw [->, thick] (0.5, 0.5) -- (0.5, 0) -- (1.3, 0);
\draw [thick] (1.3, 0.6) rectangle (2.7,1.4);
\draw [thick] (1.3, -0.4) rectangle (2.7,0.4);

\draw  (3.55,1 ) circle [radius=0.15];
\draw[shift={(3.55, 1)}] (0, -0.15) -- (0, 0.15);
\draw[shift={(3.55, 1)}] (-0.15, 0) -- ( 0.15, 0);
\draw  (3.55,0 ) circle [radius=0.15];
\draw[shift={(3.55, 0)}] (0, -0.15) -- (0, 0.15);
\draw[shift={(3.55, 0)}] (-0.15, 0) -- ( 0.15, 0);
\draw [fill] (0, 0.5) circle [radius=1pt];

\begin{scope}[shift={(1,0)}]
\draw [ thick] (2.7, 1) -- (3.5, 1) -- (3.5, 0.5);
\draw [ thick] (2.7, 0) -- (3.5, 0) -- (3.5, 0.5);
\draw [ thick] (3.5, 0.5) -- (4.0, 0.5);
\draw [fill] (4.0, 0.5) circle [radius=1pt];
\node [above] at (4.0, 0.5) {\small $\hat{\bfr}_k$};
\node [above left] at (3.5, 1) {\scriptsize $r^+_k$};
\node [below left] at (3.5, 0) {\scriptsize $r^-_k$};
\draw [fill] (7.6, 0.5) circle [radius=1pt];
\draw [thick] (4.0, 0.5) -- (4.6, 0.5);
\draw [thick] (4.6, 0) rectangle (6.7, 1);
\draw [thick] (6.7, 0.5) -- (7.6, 0.5);
\node [above] at (7.6, 0.5) {\small $\bfr_k$};
\node at (5.65, 0.5) 
{\scriptsize $\hat{\bfr}_k^T=\Lambda U \bfr_k^T$};
\node at (5.65, 1.5)
{\scriptsize
$\Lambda=$ {\tiny $\begin{bmatrix}\begin{smallmatrix} \frac{1}{1+\epsilon} & 0 \\ 0 & \frac{1}{1-\epsilon}  \end{smallmatrix} \end{bmatrix}$}
};
\end{scope}

\node [above] at (-3.2, 0.5) {\small $\bfx_k$};
\node [above] at (0, 0.5) {\small $\bfz_k$};
\node [above right] at (0.5, 1) {\scriptsize $z^+_k$};
\node [below right] at (0.5, 0) {\scriptsize $z^-_k$};

\node at (2, 0) {\footnotesize $h(D)$};
\node at (2, 1) {\footnotesize $h(D)$};

\node at (-1.4, 0.5) 
{\scriptsize $\bfz_k^T=U\bfx_k^T$};
\node at (-1.4, 1.5)
{\scriptsize
$U=\begin{bmatrix}\begin{smallmatrix} 1 & 1 \\ 1 & -1 \end{smallmatrix} \end{bmatrix}$
};

\draw [ thick] (2.1, 2.5)--(3.55, 2.5)--(3.55, 2.15);
\draw  (3.55,2 ) circle [radius=0.15];
\draw[shift={(3.55, 2)}, rotate=45] (0, -0.15) -- (0, 0.15);
\draw[shift={(3.55, 2)}, rotate=45] (-0.15, 0) -- ( 0.15, 0);
\draw [->, thick] (3.55, 1.85) -- (3.55, 1.15);
\node [left] at (3.55, 1.7) {\scriptsize $n^+_k$};
\node [right] at (3.55, 2) {\tiny $1/(1+\epsilon)$};

\draw [ thick] (2.1, -1.5)--(3.55, -1.5)--(3.55, -1.15);
\draw  (3.55,-1 ) circle [radius=0.15];
\draw[shift={(3.55, -1)}, rotate=45] (0, -0.15) -- (0, 0.15);
\draw[shift={(3.55, -1)}, rotate=45] (-0.15, 0) -- ( 0.15, 0);
\draw [->, thick] (3.55, -0.85) -- (3.55, -0.15);
\node [right] at (3.55, -1) {\tiny $1/(1-\epsilon)$};
\node [left] at (3.55, -0.7) {\scriptsize $n^-_k$};

\node [left] at (2.1, 2.5) {\scriptsize $n^a_k+n^b_k$};
\node [left] at (2.1, -1.5) {\scriptsize $n^a_k-n^b_k$};
\draw [thick] (2.7, 1) -- (3.4,1);
\draw [thick] (2.7, 0) -- (3.4,0);
\end{tikzpicture}
\caption{ A schematic of the WSSJD model. Coordinate transformations are applied in the input space ($\bfz_k^T=U\bfx_k^T$) and the output space ($\hat{\bfr}_k^T=\Lambda U \bfr_k^T$). They decompose the original 2H2T system into the sum channel (upper branch) and the subtract channel (lower branch).}
\label{fig_wssjd}
\end{figure}

Let $\bfx_k=(x^a_k, x^b_k)$ and $\bfr_k=(r^a_k,r^b_k)$ be the input and received symbols of the original system (\ref{eq_1}) and let $\bfz_k=(z^+_k, z^-_k)$ and $\hat{\bfr}_k=(r^+_k,r^-_k)$ denote the input and received symbols of the transformed system (\ref{eq_minus}). Their equivalence is visually indicated in \Figref{fig_wssjd}. By applying the coordinate transformations both in the input space and the output space, we decompose the original 2H2T system into two separate channels, both of which are modeled by channel polynomial $h(D)$. The noise components of the transformed system, $n_k^+$ and $n_k^-$, are independent, but with different noise power, $n^+_k \sim \cN(0,\frac{2\sigma^2}{(1+\epsilon)^2})$, $n^-_k \sim \cN(0, \frac{2\sigma^2}{(1-\epsilon)^2})$. 

The ML trellis of the transformed system is formed by all possible $\bfp_n=[\bfz_{n-1},\dots,\bfz_{n-\nu}]$. Notice that in this new trellis the branch labels are independent of $\epsilon$ since it only affects the noise power of the sum/subtract channel. The ML detection on this new channel is called \textbf{weighted sum subtract joint detection} (WSSJD), summarized as follows:
\begin{enumerate}
\item Calculate $r^+(D)$ and $r^-(D)$ using \Equref{eq_r}.
\item To apply VA on the WSSJD trellis, weight the branch metrics
\begin{align}
& m(\bfp_n, \bfp_{n+1}) \notag \\ & \quad =(1+\epsilon)^2(r_n^+-y_n^+)^2+(1-\epsilon)^2(r_n^--y_n^-)^2,
\label{eq_metric}
\end{align}
where $y^+_n=\sum_{i=0}^{\nu}h_iz^+_{n-i}$ and $y^-_n=\sum_{i=0}^{\nu}h_iz^-_{n-i}$ are the noiseless ISI channel outputs. Choose the path with the smallest metric and decode to the estimates $\hat{z}^+(D)$ and $\hat{z}^-(D)$.
\item Calculate $\hat{x}^a(D)$ and $\hat{x}^b(D)$ using \Equref{eq_z}.
\end{enumerate}

%\begin{figure}
%\centering
%\includegraphics[width=0.8\columnwidth]{figures/new_wssjd_trellis}
%\caption{A full WSSJD trellis for channel $h(D)=1+D$. The text next to each state lists the labels for each branch, in form of input/output}
%\label{fig_wssjdtrellis}
%\end{figure}

WSSJD has the same performance as the ML detector \cite{Fan1506}. Therefore, in the simulation we use WSSJD as a MLSE substitute for the 2H2T system, and the subset trellis is also constructed by considering the WSSJD inputs/outputs. As we will see, the coordinate transformations in WSSJD make it easier to measure the distances between symbols, which plays an important role in designing the set partition tree. Moreover, the structure of parallel channels can provide additional complexity reduction in selecting survivor paths. However, the applicability of RSSE to the standard 2H2T ML detector holds. With a little abuse of notation, when we mention the ``ML trellis'', we refer to the full ``WSSJD trellis''. For more information about WSSJD, the reader is referred to \cite{Fan1506}.

\begin{figure}
\centering
\begin{minipage}[c]{0.35\columnwidth}
\begin{tikzpicture}
\draw [thick] (0, 1) -- (2,1);
\draw [thick] (1, 0) -- (1,2);
\draw [fill] (0.3, 1) circle [radius=2pt];
\draw [fill] (1, 0.3) circle [radius=2pt];
\draw [fill] (1.7, 1) circle [radius=2pt];
\draw [fill] (1, 1.7) circle [radius=2pt];
\node [below] at (1.7, 1) {\small $(+2, 0)$};
\node [below] at (0.3, 1) {\small $(-2, 0)$};
\node [above right] at (1, 1.7) {\small $(0, +2)$};
\node [below right] at (1, 0.3) {\small $(0, -2)$};
\node [right] at (2,1) {$z^+$};
\node [above] at (1,2) {$z^-$};
\end{tikzpicture}
\end{minipage}~
\begin{minipage}[c]{0.63\columnwidth}
{
\footnotesize
\begin{tabular}{|c|c|}
\hline
$(\bfz_k, \tilde{\bfz}_k)$ & $d(\bfz_k, \tilde{\bfz}_k)$ \\
\hline \hline
$((+2, 0),(-2,0))$ & $\Delta_1^2=8(1+\epsilon)^2$\\
\hline
$((0, +2),(0,-2))$ & $\Delta_2^2=8(1-\epsilon)^2$\\
\hline
$((+2, 0),(0,+2))$ &  \\
$((+2, 0),(0,-2))$ & $\Delta_3^2=4(1+\epsilon^2)$\\
$((-2, 0),(0,+2))$ & \\
$((-2,0),(0,-2))$ & \\
\hline
\end{tabular}}
\end{minipage}
\caption{The input constellation (left) and the ESPDs (right).}
\label{fig_constel}
\end{figure}
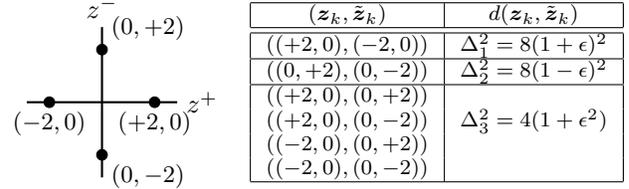
\begin{figure}
\centering
\includegraphics[width=0.82\columnwidth]{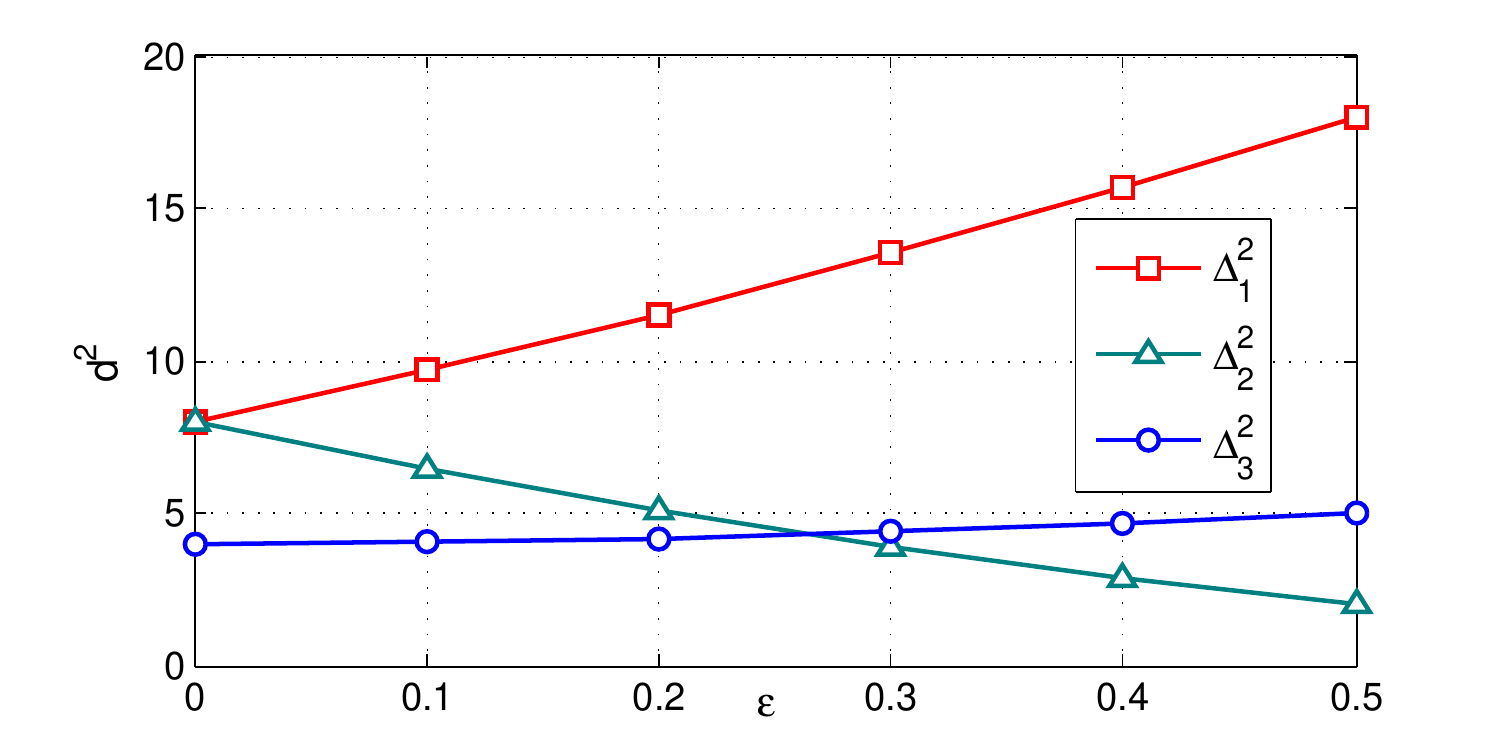}
\caption{ESPDs as functions of $\epsilon$}
\label{fig_distances}
\end{figure}

The transformations in WSSJD decompose the original 2H2T system into two parallel channels. The sum channel and the subtract channel correspond to transmitting $z^+(D)$ and $z^-(D)$ through $h(D)$, respectively. Recall that in the QAM system, the real and imaginary components of a complex symbol are also transmitted through the channel independently. Therefore, $z^+_k$ and $z^-_k$ can be treated as the real and imaginary components of a complex symbol $\bfz_k$. The only difference from the QAM system is that the sum and the subtract channels have different signal-to-noise ratios (SNRs). The sum channel is less noisy, which results in more reliable early merge than the subtract channel. Considering this dimensional asymmetry, instead of using a Euclidean distance we define the {\bfseries effective symbol pair distance} (ESPD)
\begin{align}
& d^2(\bfz_k, \tilde{\bfz}_k) \notag \\ & \quad=\frac{(1+\epsilon)^2}{2}(z^+_k-\tilde{z}^+_k)^2+\frac{(1-\epsilon)^2}{2}(z^-_k-\tilde{z}^-_k)^2. \label{eq_10}
\end{align}

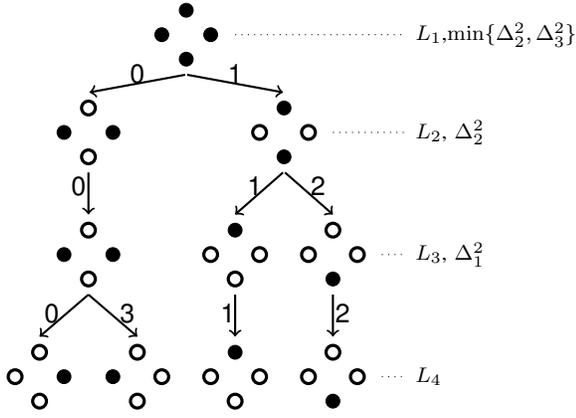
\begin{figure} % % % % set partition figure
\centering
\begin{tikzpicture} [scale=0.65]
%\node[above] at (4, 9) {Set partitioning};
%\draw [help lines] (0,0) grid (8,8);
%\draw [thick] (0.2, 1) -- (1.8, 1);
%\draw [thick] (1, 0.2) -- (1, 1.8);
\draw [very thick] (0.5, 1) circle [radius=4pt];
\draw [fill] (1.5, 1) circle [radius=4pt];
\draw [very thick] (1, 0.5) circle [radius=4pt];
\draw [very thick] (1, 1.5) circle [radius=4pt];
%\draw [red, thick] (1.5, 1) circle [radius = 3.2pt];
\begin{scope}[shift={(2,0)}]
%	\draw [thick] (0.2, 1) -- (1.8, 1);
%	\draw [thick] (1, 0.2) -- (1, 1.8);
	\draw [fill] (0.5, 1) circle [radius=4pt];
	\draw [very thick] (1.5, 1) circle [radius=4pt];
	\draw [very thick] (1, 0.5) circle [radius=4pt];
	\draw [very thick] (1, 1.5) circle [radius=4pt];
%	\draw [red, thick] (0.5, 1) circle [radius = 3.2pt];
\end{scope}
\begin{scope}[shift={(4,0)}]
%	\draw [thick] (0.2, 1) -- (1.8, 1);
%	\draw [thick] (1, 0.2) -- (1, 1.8);
	\draw [very thick] (0.5, 1) circle [radius=4pt];
	\draw [very thick] (1.5, 1) circle [radius=4pt];
	\draw [very thick] (1, 0.5) circle [radius=4pt];
	\draw [fill] (1, 1.5) circle [radius=4pt];
	%\draw [red, thick] (1, 1.5) circle [radius = 3.2pt];
\end{scope}
\begin{scope}[shift={(6,0)}]
%	\draw [thick] (0.2, 1) -- (1.8, 1);
%	\draw [thick] (1, 0.2) -- (1, 1.8);
	\draw [very thick] (0.5, 1) circle [radius=4pt];
	\draw [very thick] (1.5, 1) circle [radius=4pt];
	\draw [fill] (1, 0.5) circle [radius=4pt];
	\draw [very thick] (1, 1.5) circle [radius=4pt];
%	\draw [red, thick] (1, .5) circle [radius = 3.2pt];
\end{scope}

\begin{scope}[shift={(1,2.5)}]
%	\draw [thick] (0.2, 1) -- (1.8, 1);
%	\draw [thick] (1, 0.2) -- (1, 1.8);
	\draw [fill] (0.5, 1) circle [radius=4pt];
	\draw [fill] (1.5, 1) circle [radius=4pt];

	\draw [very thick] (1, 0.5) circle [radius=4pt];
	\draw [very thick] (1, 1.5) circle [radius=4pt];
%	\draw [red, thick] (0.5, 1) circle [radius = 3.2pt];
%	\draw [red, thick] (1.5, 1) circle [radius = 3.2pt];
\end{scope}
\begin{scope}[shift={(4,2.5)}]
%	\draw [thick] (0.2, 1) -- (1.8, 1);
%	\draw [thick] (1, 0.2) -- (1, 1.8);
	\draw [very thick] (0.5, 1) circle [radius=4pt];
	\draw [very thick] (1.5, 1) circle [radius=4pt];
	\draw [very thick] (1, 0.5) circle [radius=4pt];
	\draw [fill] (1, 1.5) circle [radius=4pt];
%	\draw [red, thick] (1, 1.5) circle [radius = 3.2pt];
\end{scope}
\begin{scope}[shift={(6,2.5)}]
%	\draw [thick] (0.2, 1) -- (1.8, 1);
%	\draw [thick] (1, 0.2) -- (1, 1.8);
	\draw [very thick] (0.5, 1) circle [radius=4pt];
	\draw [very thick] (1.5, 1) circle [radius=4pt];
	\draw [fill] (1, 0.5) circle [radius=4pt];
	\draw [very thick] (1, 1.5) circle [radius=4pt];
%	\draw [red, thick] (1, .5) circle [radius = 3.2pt];
\end{scope}

\begin{scope}[shift={(1,5)}]
%	\draw [thick] (0.2, 1) -- (1.8, 1);
%	\draw [thick] (1, 0.2) -- (1, 1.8);
	\draw [fill] (0.5, 1) circle [radius=4pt];
	\draw [fill] (1.5, 1) circle [radius=4pt];
	\draw [very thick] (1, 0.5) circle [radius=4pt];
	\draw [very thick] (1, 1.5) circle [radius=4pt];
%	\draw [red, thick] (0.5, 1) circle [radius = 3.2pt];
%	\draw [red, thick] (1.5, 1) circle [radius = 3.2pt];
\end{scope}
\begin{scope}[shift={(5,5)}]
%	\draw [thick] (0.2, 1) -- (1.8, 1);
%	\draw [thick] (1, 0.2) -- (1, 1.8);
	\draw [very thick] (0.5, 1) circle [radius=4pt];
	\draw [very thick] (1.5, 1) circle [radius=4pt];
	\draw [fill] (1, 0.5) circle [radius=4pt];
	\draw [fill] (1, 1.5) circle [radius=4pt];

\end{scope}
\begin{scope}[shift={(3,7)}]
%	\draw [thick] (0.2, 1) -- (1.8, 1);
%	\draw [thick] (1, 0.2) -- (1, 1.8);
	\draw [fill] (0.5, 1) circle [radius=4pt];
	\draw [fill] (1.5, 1) circle [radius=4pt];
	\draw [fill] (1, 0.5) circle [radius=4pt];
	\draw [fill] (1, 1.5) circle [radius=4pt];
\end{scope}

\draw[->, thick] (3.98,7.18) -- (2.02, 6.82);
\node at (3, 7.2) {\sffamily 0};
\draw[->,thick] (4.02,7.18) -- (5.98, 6.82);
\node at (5, 7.2) {\sffamily 1};
\draw[->,thick] (2,5.18) -- (2, 4.32);
\node at (1.8, 4.9) {\sffamily 0};
\draw[->,thick] (5.98,5.18) -- (5, 4.32);
\node at (5.4, 4.9) {\sffamily 1};
\draw[->,thick] (6.02,5.18) -- (7, 4.32);
\node at (6.7, 4.9) {\sffamily 2};
\draw[->,thick] (2,2.68) -- (1, 1.82);
\node at (1.25, 2.3) {\sffamily 0};
\draw[->,thick] (2,2.68) -- (3, 1.82);
\node at (2.8, 2.3) {\sffamily 3};
\draw[->,thick] (5,2.68) -- (5, 1.82);
\node at (4.85, 2.3) {\sffamily 1};
\draw[->,thick] (7,2.68) -- (7, 1.82);
\node at (7.2,2.3) {\sffamily 2};

\draw [dotted] (5,8) -- (8.5,8);
\draw [dotted] (7,6) -- (8.5,6);
\draw [dotted] (8,3.5) -- (8.5,3.5);
\draw [dotted] (8,1) -- (8.5,1);
\node [align=center,right] at (8.5,8) {\footnotesize $L_1$,\footnotesize $\min \{ \Delta_2^2, \Delta_3^2\}$};
\node [right] at (8.5,6) {\footnotesize $L_2$, $\Delta_2^2$};
\node [right] at (8.5,3.5) {\footnotesize $L_3$, $\Delta_1^2$};
\node [right] at (8.5,1) {\footnotesize $L_4$};
\end{tikzpicture}
\caption{The modified set partition tree. This tree contains $4$ levels, $\{L_1, L_2, L_3, L_4\}$, each of which is a set partition of the WSSJD input constellation. The minimum ESPD on each level is specified on the right side. The number associated with each branch is the index of the subset in the corresponding set partition.}
\label{fig_tree}
\end{figure}

The input constellation and the ESPDs between different pairs of symbols are shown in \Figref{fig_constel}. Notice that ESPDs can change with respect to $\epsilon$, as shown in \Figref{fig_distances}. Therefore even with the same subset trellis configuration, the RSSE performs differently at different ITI levels.

The set partition tree designed for 2H2T is constructed by maximizing the minimum intrasubset ESPD on each level, which results in an unbalanced tree shown in \Figref{fig_tree}. Compared to the Ungerboeck set partition tree, the additional level $L_3$ comes from the asymmetric distance measure in the $z^+$ and $z^-$ dimensions, and it provides more flexibility in choosing set partitions, which leads to a better performance/complexity tradeoff.

\begin{figure} % % subset trellis for PR2 channel
\centering
\begin{tikzpicture}[scale = 0.55]

\draw[fill] (0,0) circle [radius=1.5pt]  node[left] {{\scriptsize $[3,1]\ $}};
\draw[fill] (0,1) circle [radius=1.5pt] node[left]  {{\scriptsize $[3,0]\ $}};
\draw[fill] (0,2) circle [radius=1.5pt] node[left]  {{\scriptsize $[2,1]\ $}};
\draw[fill] (0,3) circle [radius=1.5pt]  node[left] {{\scriptsize $[2,0]\ $}};
\draw[fill] (0,4) circle [radius=1.5pt] node[left]  {{\scriptsize $[1,1]\ $}};
\draw[fill] (0,5) circle [radius=1.5pt] node[left]  {{\scriptsize $[1,0]\ $}};
\draw[fill] (0,6) circle [radius=1.5pt] node[left]  {{\scriptsize $[0,1]\ $}};
\draw[fill] (0,7) circle [radius=1.5pt] node[left]  {{\scriptsize $[0,0]\ $}};
\node[above left] at (-0.2,7.2) {\scriptsize $\bfs_{n}$};
\node[above] at (-3.5,7.2) {\scriptsize $a_n(1)$};
\node at (-3.5,7) {\scriptsize $\{0,1,2,3\}$};
\node at (-3.5,6) {\scriptsize $\{0,1,2,3\}$};
\node at (-3.5,5) {\scriptsize $\{0,1,2,3\}$};
\node at (-3.5,4) {\scriptsize $\{0,1,2,3\}$};
\node at (-3.5,3) {\scriptsize $\{0,1,2,3\}$};
\node at (-3.5,2) {\scriptsize $\{0,1,2,3\}$};
\node at (-3.5,1) {\scriptsize $\{0,1,2,3\}$};
\node at (-3.5,0) {\scriptsize $\{0,1,2,3\}$};

\begin{scope}[shift={(5,0)}]
\draw[fill] (0,0) circle [radius=1.5pt]  node[right] {{\scriptsize $\ [3,1]$}};
\draw[fill] (0,1) circle [radius=1.5pt] node[right]  {{\scriptsize $\ [3,0]$}};
\draw[fill] (0,2) circle [radius=1.5pt] node[right]  {{\scriptsize $\ [2,1]$}};
\draw[fill] (0,3) circle [radius=1.5pt]  node[right] {{\scriptsize $\ [2,0]$}};
\draw[fill] (0,4) circle [radius=1.5pt] node[right]  {{\scriptsize $\ [1,1]$}};
\draw[fill] (0,5) circle [radius=1.5pt] node[right]  {{\scriptsize $\ [1,0]$}};
\draw[fill] (0,6) circle [radius=1.5pt] node[right]  {{\scriptsize $\ [0,1]$}};
\draw[fill] (0,7) circle [radius=1.5pt] node[right]  {{\scriptsize $\ [0,0]$}};
\node[above right] at (0,7.2) {\scriptsize $\bfs_{n+1}$};
\end{scope}

\draw (0,0) to (5,1);
\draw (0,0) to (5,3);
\draw (0,0) to (5,5);
\draw (0,0) to (5,7);

\draw (0,1) to (5,1);
\draw (0,1) to (5,3);
\draw (0,1) to (5,5);
\draw (0,1) to (5,7);

\draw (0,6) to (5,1);
\draw (0,6) to (5,3);
\draw (0,6) to (5,5);
\draw (0,6) to (5,7);

\draw (0,7) to (5,1);
\draw (0,7) to (5,3);
\draw (0,7) to (5,5);
\draw (0,7) to (5,7);

\draw (0,2) to (5,0);
\draw (0,2) to (5,2);
\draw (0,2) to (5,4);
\draw (0,2) to (5,6);

\draw (0,3) to (5,0);
\draw (0,3) to (5,2);
\draw (0,3) to (5,4);
\draw (0,3) to (5,6);

\draw (0,4) to (5,0);
\draw (0,4) to (5,2);
\draw (0,4) to (5,4);
\draw (0,4) to (5,6);

\draw (0,5) to (5,0);
\draw (0,5) to (5,2);
\draw (0,5) to (5,4);
\draw (0,5) to (5,6);
\end{tikzpicture}
\caption{Subset trellis with configuration [4,2] on memory-2 channel.}
\label{fig_pr2}
\end{figure}
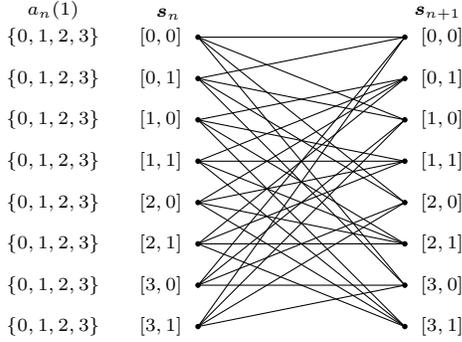

When constructing the subset trellis, we choose $\Omega(k)$ from the levels of the set partition tree for each $k=1, \cdots, \nu$, and to guarantee a well-defined trellis structure, $\Omega(k)$ should always be at the same as or higher level than $\Omega(k-1)$. As an example, \Figref{fig_pr2} shows an eight-state subset trellis for a memory-2 channel with configuration $[4,2]$, i.e., $\Omega(1)$ is chosen to be $L_4$, and $\Omega(2)$ is chosen to be $L_2$. In this case, the incoming symbols start to merge at the second stage. 

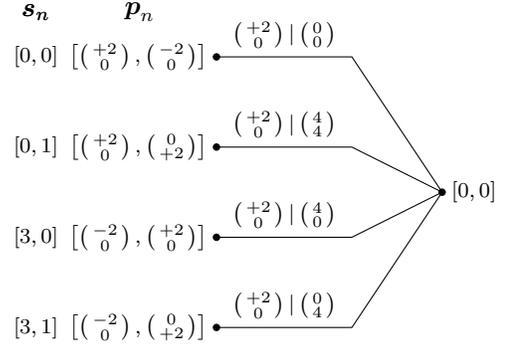
\begin{figure}
\centering
\begin{tikzpicture} [scale = 0.6]
\draw[fill] (0,6) circle [radius=2pt]  node[left] {{\footnotesize $\left[\left(\begin{smallmatrix}
+2 \\ 0
\end{smallmatrix}\right), \left(\begin{smallmatrix}
-2 \\ 0
\end{smallmatrix}\right) \right]$}};
\draw[fill] (0,4) circle [radius=2pt]  node[left] {{\footnotesize $\left[\left(\begin{smallmatrix}
+2 \\ 0
\end{smallmatrix}\right), \left(\begin{smallmatrix}
0 \\ +2
\end{smallmatrix}\right) \right]$}};
\draw[fill] (0,2) circle [radius=2pt]  node[left] {{\footnotesize $\left[\left(\begin{smallmatrix}
-2 \\ 0
\end{smallmatrix}\right), \left(\begin{smallmatrix}
+2 \\ 0
\end{smallmatrix}\right) \right]$}};
\draw[fill] (0,0) circle [radius=2pt]  node[left] {{\footnotesize $\left[\left(\begin{smallmatrix}
-2 \\ 0
\end{smallmatrix}\right), \left(\begin{smallmatrix}
0 \\ +2
\end{smallmatrix}\right) \right]$}};
\draw (0,6) -- (3,6) -- (5, 3);
\draw (0,4) -- (3,4) -- (5, 3);
\draw (0,2) -- (3,2) -- (5, 3);
\draw (0,0) -- (3,0) -- (5, 3);

\node at (-4, 6) {\footnotesize $[0,0]$};
\node at (-4, 4) {\footnotesize $[0,1]$};
\node at (-4, 2) {\footnotesize $[3,0]$};
\node at (-4, 0) {\footnotesize $[3,1]$};

\node at (-4, 7) {$\bfs_n$};
\node at (-1.7, 7) {$\bfp_n$};

\node [above] at (1.5, 6) {{\footnotesize $\left(\begin{smallmatrix}
+2 \\ 0
\end{smallmatrix}\right) | \left(\begin{smallmatrix}
0 \\ 0
\end{smallmatrix}\right)$}};
\node [above] at (1.5, 4) {{\footnotesize $\left(\begin{smallmatrix}
+2 \\ 0
\end{smallmatrix}\right) | \left(\begin{smallmatrix}
4 \\ 4
\end{smallmatrix}\right)$}};
\node [above] at (1.5, 2) {{\footnotesize $\left(\begin{smallmatrix}
+2 \\ 0
\end{smallmatrix}\right) | \left(\begin{smallmatrix}
4 \\ 0
\end{smallmatrix}\right)$}};
\node [above] at (1.5, 0) {{\footnotesize $\left(\begin{smallmatrix}
+2 \\ 0
\end{smallmatrix}\right) | \left(\begin{smallmatrix}
0 \\ 4
\end{smallmatrix}\right)$}};

\draw[fill] (5,3) circle [radius=2pt]  node[right] {{\footnotesize $[0,0]$}};
\node at (-4, 7) {$\bfs_n$};
\end{tikzpicture}
\caption{An illustration of detection on subset trellis. Consider a subset trellis $[4,2]$ constructed for PR2 channel $1+2D+D^2$. The leftmost column lists the subset state $\bfs_n$, while the column next to it lists the corresponding survivor ML states $\bfp_n$. The labels on the branches are formed as $\mathcal{L}_{\text{in}}|\mathcal{L}_{\text{out}}$. The branches terminate at subset state $\bfs_{n+1}=[0,0]$.}
\label{fig_selection}
\end{figure}

Although one subset state is a group of several ML trellis states, only one ML state can survive inside each subset state at one time slot. Consider the example shown in 
Fig. \ref{fig_selection} and assume $\epsilon=0.1$. The labels next to the states list the subset states and their current survivor ML states. The labels on the branches are the channel input/output. Notice that the outputs are calculated based on the current $\bfp_n$. A look-up table can be stored to facilitate the process of finding the corresponding output labels once the system decides the survivor ML states. Assume the received signals are  $\hat{\bfr}=(r^+,r^-)=(5,3)$. The metric comparison shows that the subset state $[0,1]$ with survivor ML state $\left[\left(\begin{smallmatrix}
+2 \\ 0
\end{smallmatrix}\right), \left(\begin{smallmatrix}
0 \\ +2
\end{smallmatrix}\right) \right]$ gives the smallest path metric, so the survivor ML state of $\bfs_{n+1}=[0,0]$ can be decided and updated to be $\left[\left(\begin{smallmatrix}
+2 \\ 0
\end{smallmatrix}\right), \left(\begin{smallmatrix}
+2 \\ 0
\end{smallmatrix}\right) \right]$, which will be used in the next time slot.

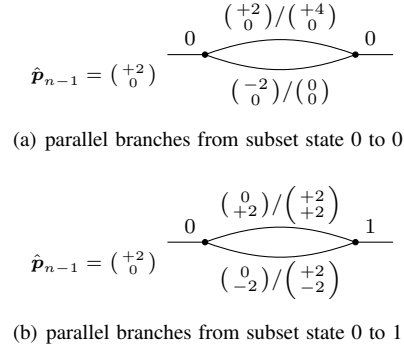
\begin{figure}
\centering
\subfigure[parallel branches from subset state 0 to 0]{
\begin{tikzpicture}[scale=0.5]
\draw[fill] (0,0) circle [radius=2pt] ;
\draw[fill] (4,0) circle [radius=2pt];
\node[above right] at (4,0) {{\footnotesize $0$}};
\node[above left] at (0,0) {{\footnotesize $0$}};
%\node[below left, align=left] at (0,0) {\scriptsize $\hat{\bfp}_{n-1}$ \\ \scriptsize $=\begin{pmatrix}
%\begin{smallmatrix}
%+2 \\ 0
%\end{smallmatrix}
%\end{pmatrix}$};
\node[below left, align=left] at (-1,0) {\scriptsize $\hat{\bfp}_{n-1}=\begin{pmatrix}
\begin{smallmatrix}
+2 \\ 0
\end{smallmatrix}
\end{pmatrix}$};

\draw (0,0) to [out=20, in=160] (4,0);
\draw (0,0) to [out=-20, in=-160] (4,0);
\draw (0,0) to  (-1,0);
\draw (4,0) to  (5,0);
\node at (2,1) {\footnotesize $\begin{pmatrix}
\begin{smallmatrix}
+2 \\ 0
\end{smallmatrix}
\end{pmatrix} \! /\! \begin{pmatrix}
\begin{smallmatrix}
+4 \\ 0
\end{smallmatrix}
\end{pmatrix}$ };

\node at (2,-1) {\footnotesize $\begin{pmatrix}
\begin{smallmatrix}
-2 \\ 0
\end{smallmatrix}
\end{pmatrix} \! /\! \begin{pmatrix}
\begin{smallmatrix}
0 \\ 0
\end{smallmatrix}
\end{pmatrix}$ };

\end{tikzpicture}
\label{sub_0to0}
}

\subfigure[parallel branches from subset state 0 to 1]{
\begin{tikzpicture}[scale=0.5]
\draw[fill] (0,0) circle [radius=2pt] ;
\draw[fill] (4,0) circle [radius=2pt];
\node[above right] at (4,0) {{\footnotesize $1$}};
\node[above left] at (0,0) {{\footnotesize $0$}};
\node[below left, align=left] at (-1,0) {\scriptsize $\hat{\bfp}_{n-1}=\begin{pmatrix}
\begin{smallmatrix}
+2 \\ 0
\end{smallmatrix}
\end{pmatrix}$};

\draw (0,0) to [out=20, in=160] (4,0);
\draw (0,0) to [out=-20, in=-160] (4,0);
\draw (0,0) to  (-1,0);
\draw (4,0) to  (5,0);
\node at (2,1) {\footnotesize $\begin{pmatrix}
\begin{smallmatrix}
0 \\ +2
\end{smallmatrix}
\end{pmatrix} \! /\! \begin{pmatrix}
\begin{smallmatrix}
+2 \\ +2
\end{smallmatrix}
\end{pmatrix}$ };

\node at (2,-1) {\footnotesize $\begin{pmatrix}
\begin{smallmatrix}
0 \\ -2
\end{smallmatrix}
\end{pmatrix} \! /\! \begin{pmatrix}
\begin{smallmatrix}
+2 \\ -2
\end{smallmatrix}
\end{pmatrix}$ };
\end{tikzpicture}
\label{sub_0to1}
}
\caption{Sample parallel branches for subset trellis with 2 states constructed for channel $1+D$}
\label{fig_pbranch}
\end{figure}
%\begin{figure}
%\begin{tikzpicture}[scale=0.6]
% % % graph with explicit double edges
% \node at (-0.2,4.8) {$\bfs_{n-1}$};
% \node at (5.2,4.8) {$\bfs_{n}$};
% \begin{scope}[shift={(-5,0)}]
% \node at (0,4.9) {$\hat{\bfp}_{n-1}$};
% \node at (0,4) {$(+2,0)$};
% \node at (0,0) {$(0,-2)$};
% \end{scope}
%
%%  \node at (5.2,4.5) {$\bfs_{n}$};
%\draw[fill] (0,0) circle [radius=1.5pt]  node[left] {{\footnotesize $1$}};
%\draw[fill] (0,4) circle [radius=1.5pt] node[left] {{\footnotesize $0$} };
%\draw[fill] (5,0) circle [radius=1.5pt]  node[right] {{\footnotesize $1$}};
%\draw[fill] (5,4) circle [radius=1.5pt] node[right] {{\footnotesize $0 $} } ;
%\draw (0,0) to [out=15, in=165] (5,0);
%\draw (0,0) to [out=-15, in=-165] (5,0);
%\draw (0,4) to [out=15, in=165] (5,4);
%\draw (0,4) to [out=-15, in=-165] (5,4);
%\draw (0,0) to [out=55, in=-155] (5,4);
%\draw (0,0) to [out=25, in=-125] (5,4);
%\draw (0,4) to [out=-55, in=155] (5,0);
%\draw (0,4) to [out=-25, in=125] (5,0);
%\end{tikzpicture}
%\end{figure}

For a configuration with $J_1<4$, $a_{n}(1)$ may contain more than one input symbol. % The resulting subset trellis will contain parallel branches. The example of trellis with parallel branches is shown in Figs. \ref{fig_dicode} for $1+D$ channel. The full WSSJD trellis has 4 states as shown in Fig. \ref{fig_wssjdtrellis}.
For instance, consider the $1+D$ channel. Choosing $\Omega(1)$ to be $L_2$ or $L_3$ results in a subset trellis with 2 states or 3 states, respectively. For the subset trellis with parallel branches, a pre-selection between the parallel branches is needed during the detection. Due to the symmetric property of WSSJD trellis labels, this pre-selection can be done without explicitly calculating the branch metric, if $J_1>1$. For instance, consider the two scenarios illustrated in Figs. \ref{fig_pbranch}. In both cases, the survivor ML state at the starting stage is assumed to be $\hat{\bfp}_{n-1}=(+2,0)$. The input and output labels are marked on the branches. In \Figref{fig_pbranch}\subref{sub_0to0}, both the input symbols $(+2,0)$ and $(-2,0)$ lead the paths to subset state $0$. Notice that the input symbols $(+2,0)$ and $(-2,0)$ have the same value in the $z^-$ dimension, producing the same output on the subtract channel. Instead of calculating metrics from \Equref{eq_metric}, the pre-selection performs a thresholding on the sum channel output and make the decision. In this example, the threshold is $+2$, obtained by averaging $+4$ and $0$. If $r_n^+>+2$, the strategy is to pick $(+2,0)$ as the survived symbol, while for the case $r_n^+<+2$, $(-2,0)$ should be the survivor Similarly for another case shown in \Figref{fig_pbranch}\subref{sub_0to1}, the thresholding is conducted on the subtract channel output, since the two input symbols produce the same output in the sum channel. By comparing $r_n^-$ with the threshold $0$, the detector picks $(0, +2)$ if $r_n^->0$, or $(0, -2)$ if $r_n^-<0$. This symmetry property renders the WSSJD formulation preferable over the traditional ML detector.

\section{Simulation Results}
\label{sec_simulation}
We examine the RSSE performance on various types of channels at different ITI levels. 
The SNR is defined as
\begin{align}
\text{SNR(dB)}=10\log\frac{\|h(D)\|^2}{2\sigma^2 }
\end{align}

\begin{figure}
\centering
\hspace{-2pt}
\subfigure[$\epsilon=0.1$]{
\includegraphics[width=0.5\columnwidth]{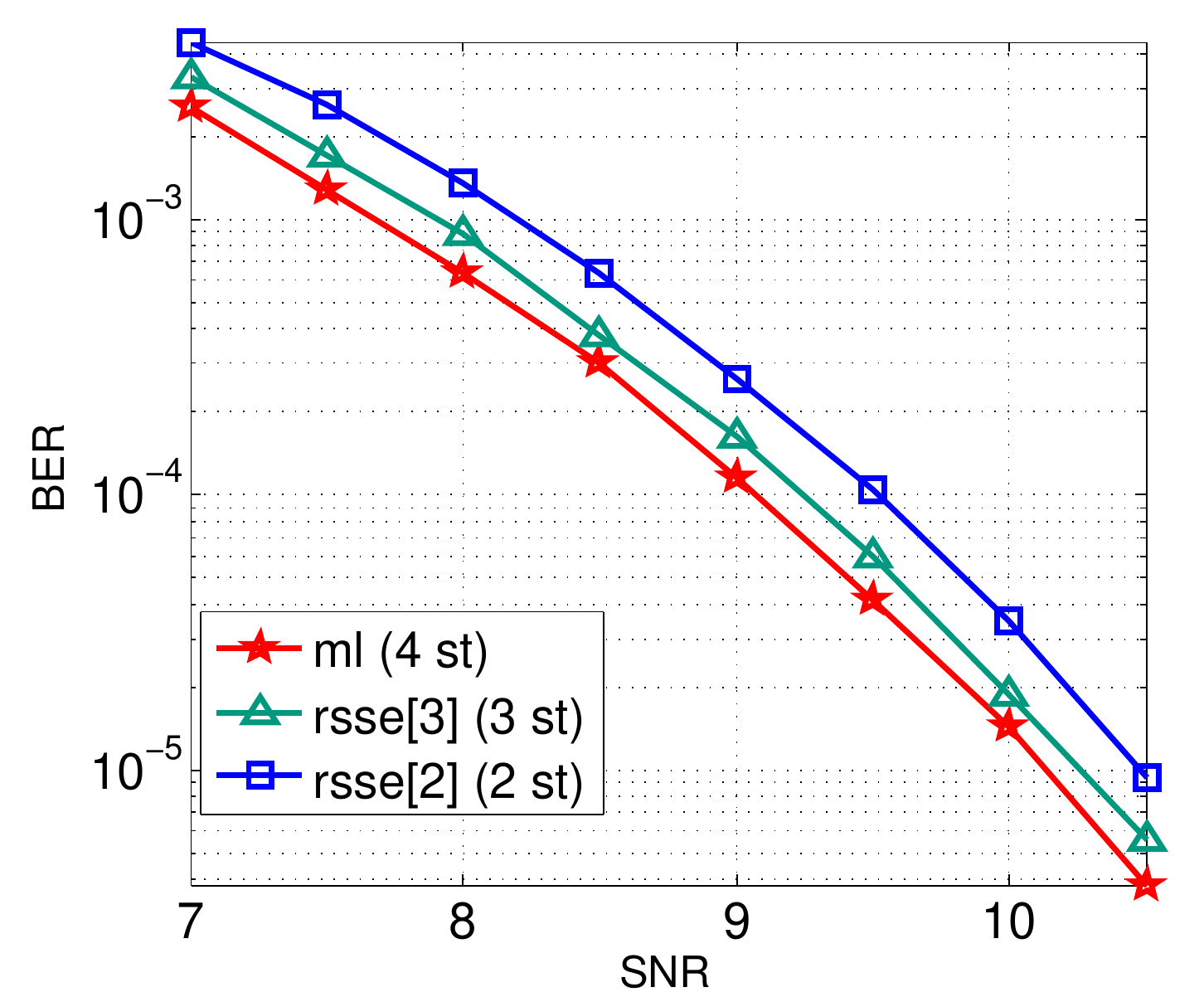}
\label{sub_dicode_1}
}~
\hspace{-12pt}
\subfigure[$\epsilon=0.3$]{
\includegraphics[width=0.5\columnwidth]{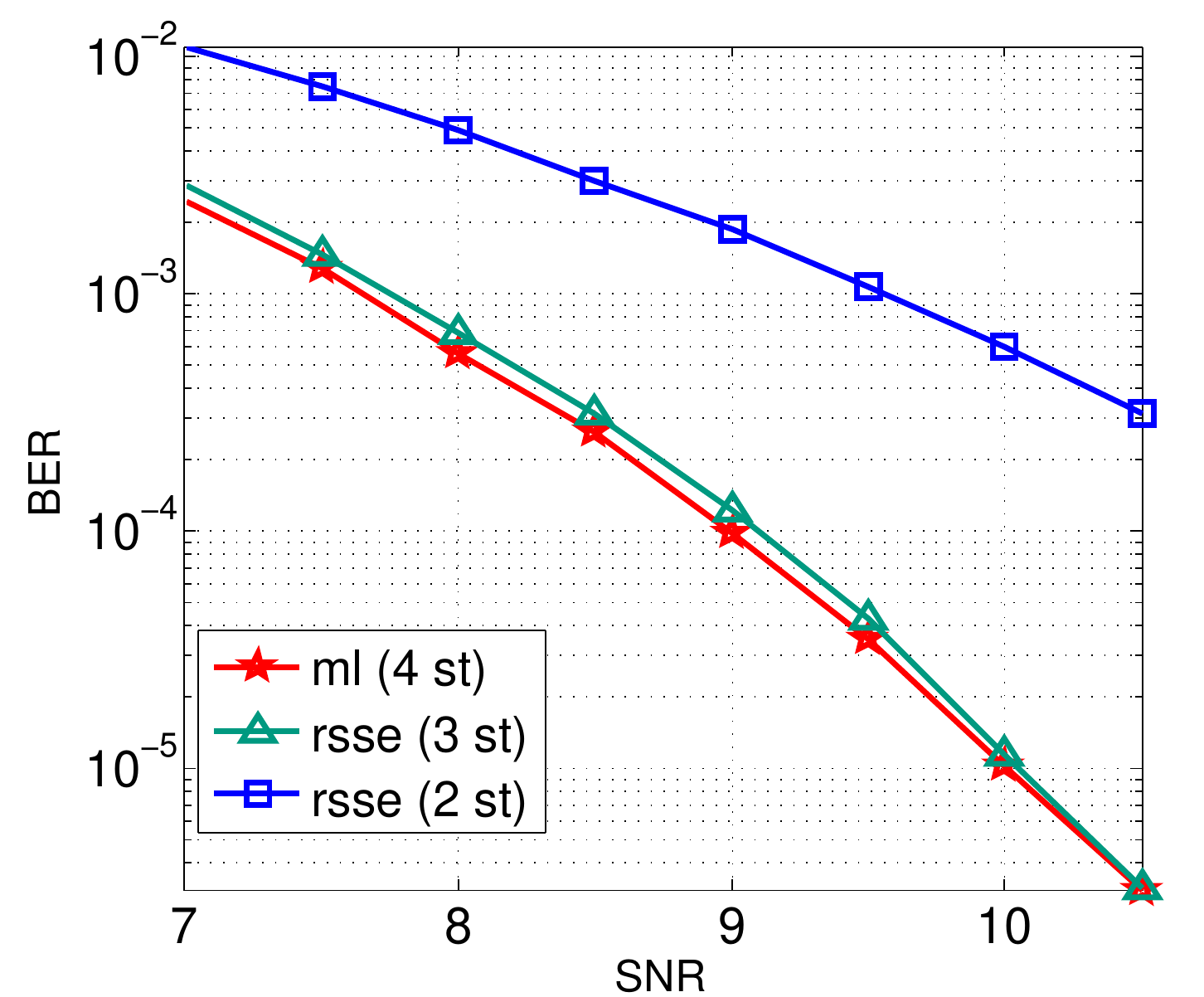}
\label{sub_dicode_3}
}
\caption{Performance comparison between RSSE and ML detector on dicide channel at different ITI levels.}

\label{fig_dicode_result}
\end{figure}

\begin{figure}
\centering
\hspace{-2pt}
\subfigure[$\epsilon=0.1$]{
\includegraphics[width=0.5\columnwidth]{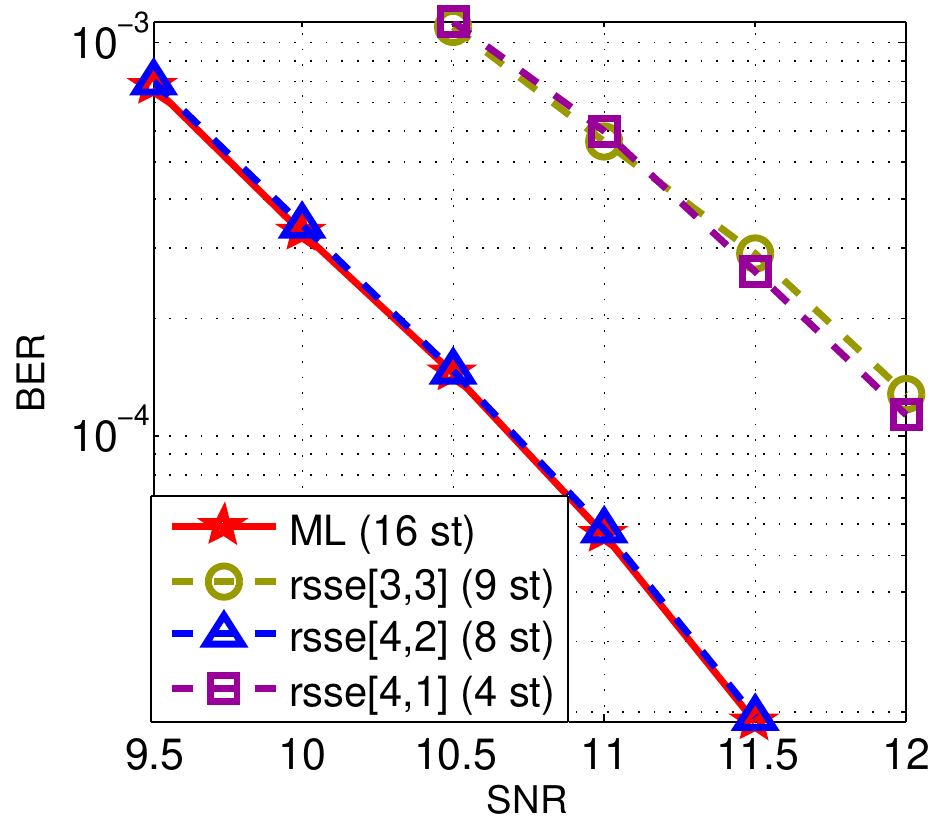}
\label{sub_pr2_1}
}~
\hspace{-12pt}
\subfigure[$\epsilon=0.3$]{
\includegraphics[width=0.5\columnwidth]{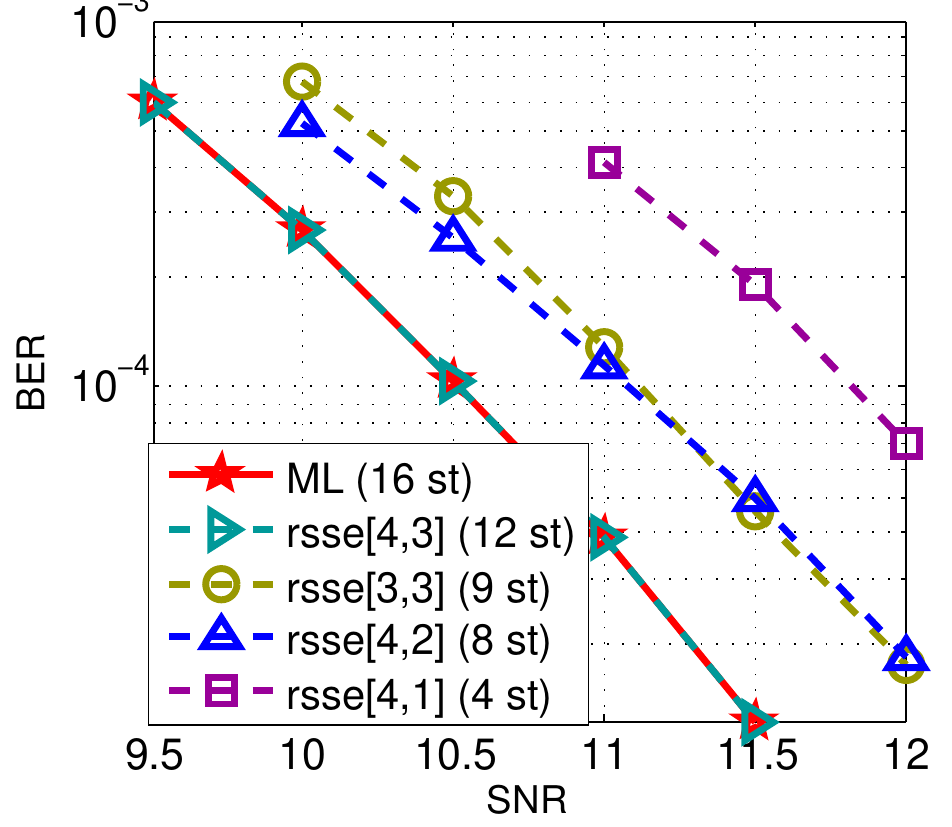}
\label{sub_pr2_3}
}
\caption{Performance comparison between RSSE and ML detector on PR2 channel at different ITI levels. The legend shows the RSSE subset trellis configuration and the corresponding number of trellis states.}

\label{fig_pr2_result}
\end{figure}

\begin{figure}
\centering
\hspace{-2pt}
\subfigure[$\epsilon=0.1$]{
\includegraphics[width=0.5\columnwidth]{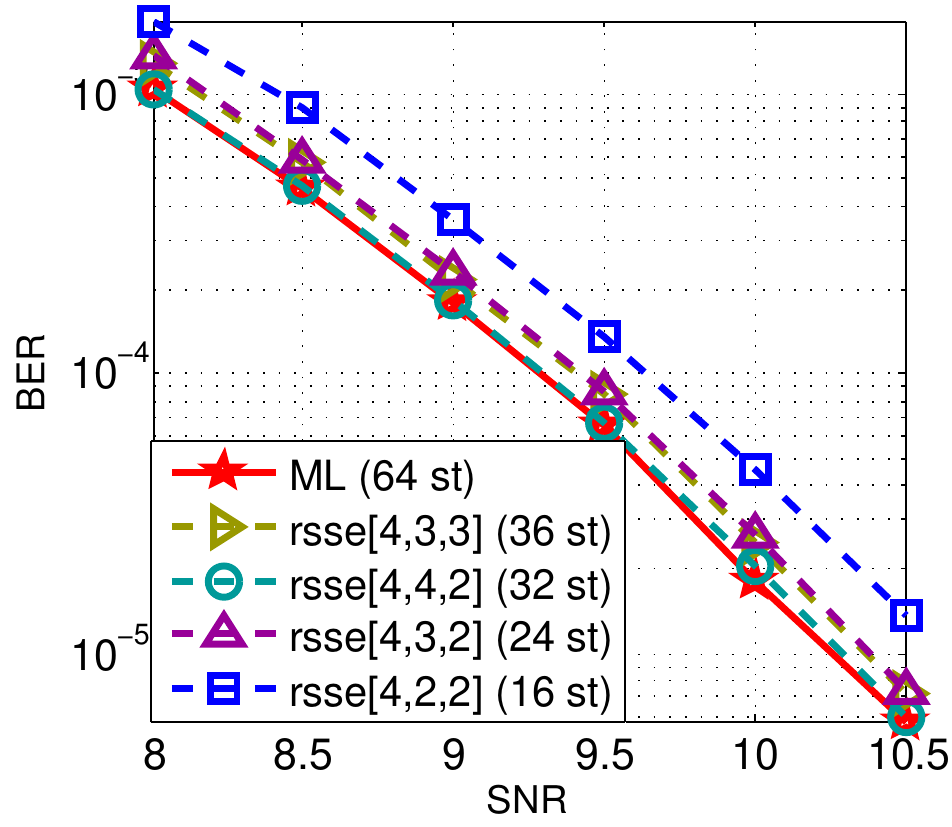}
\label{sub_epr4_1}
}~
\hspace{-12pt}
\subfigure[$\epsilon=0.3$]{
\includegraphics[width=0.5\columnwidth]{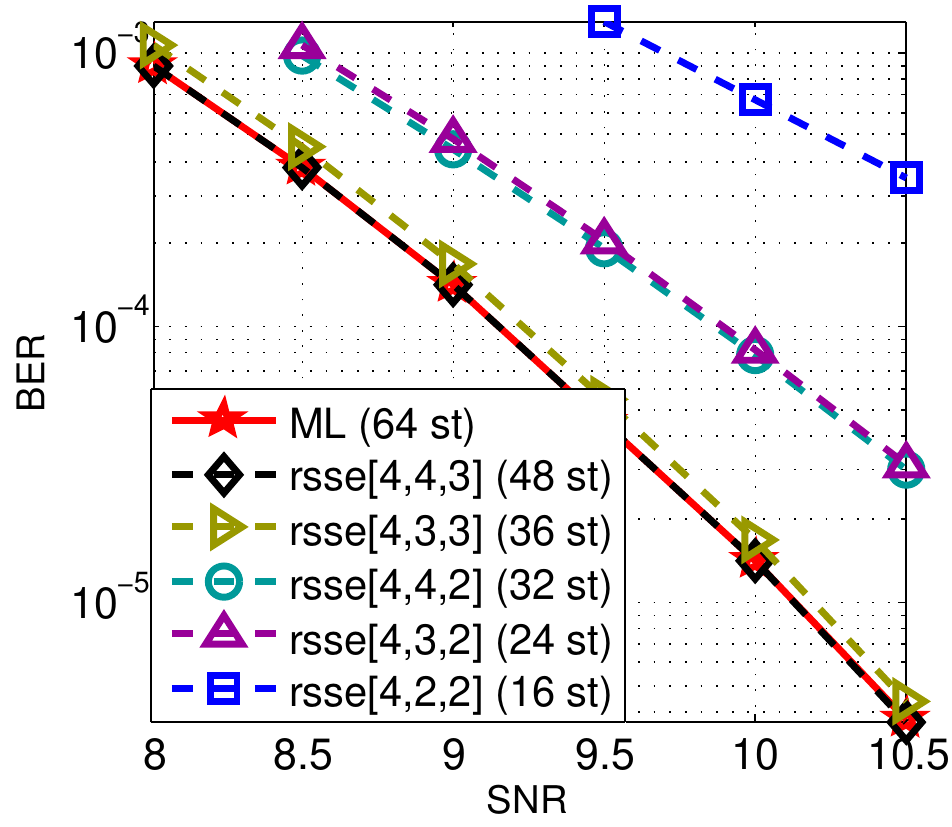}
\label{sub_epr4_3}
}
\caption{Performance comparison between RSSE and ML detector on EPR4 channel at different ITI levels. The legend shows the RSSE subset trellis configuration and the corresponding number of trellis states.}
%The vector behind ``rsse" in the legend of each subfigure represents the subset trellis configuration. The total number of trellis states is also indicated next to the configuration.}
\label{fig_epr4_channels}
\end{figure}
\vspace{-10pt}

\subsection{Dicode Channel}
This simple example helps us understand how the pre-selection between parallel branches affects the system. Although early-merging happens at every time step, its doesn't degrade the performance heavily. From Figs. \ref{fig_dicode_result} we see that the performance loss of the 3-state subset trellis is within 0.1dB. The 2 state RSSE performs better under the lower ITI level. 
 
\subsection{Channel with Higher Memory}
Higher channel memory provides more flexibility to construct the subset trellis, based on different requirements of performance/complexity tradeoff. For the EPR4 channel $h(D)=1+D-D^2-D^3$, we apply RSSE to several subset trellises with different complexities. The resulting bit error rate (BER) vs. SNR performance at different ITI levels is plotted in \Figref{fig_epr4_channels}. The comparison between Figs. \ref{sub_epr4_1} and Figs. \ref{sub_epr4_3} shows that even using the same subset trellis, RSSE performs differently under different ITI levels, and its performance correlates with the minimum intrasubset ESPDs of the set partitions configured in the subset trellis. At a low ITI level ($\epsilon=0.1$), the performance of RSSE on subset trellis $[4,4,2]$ coincides with that of the ML detector. The BER curves of $[4,3,3]$ and $[4,3,2]$ overlap, and are both within 0.1dB from the ML curve. Subset trellis $[4,2,2]$ further reduces the number of states to 16, but incurs a 0.3dB loss. When the ITI level becomes higher ($\epsilon=0.3$), the subset trellis $[4,4,2]$ cannot provide reliable early path merging because the minimum intrasubset ESPD $\Delta_2^2$ in $\Omega(3)=L_2$ is substantially reduced. However, a less aggressive construction using configuration $[4,4,3]$ achieves near-ML performance. The decrease in $\Delta_2^2$ at this ITI level also degrades the performance of RSSE$[4,2,2]$ and $[4,3,2]$. Their BER curves overlap in Fig. \ref{sub_epr4_3}. In contrast, the increase of $\Delta_1^2$ brings $[4,3,3]$ closer to the ML performance, compared to the case $\epsilon=0.1$.

Similarly, for the PR2 channel $h(D)=1+2D+D^2$, the simulation results in Figs. \ref{fig_pr2_result} show that the BER curve of RSSE$[4,2]$ essentially overlaps with that of the ML detector at $\epsilon=0.1$. For $\epsilon=0.3$, RSSE$[4,3]$ essentially achieves ML performance.

\begin{table}
\centering
\begin{tabular}{ |l |l|l|l|l|}
\hline
\multirow{2}{*}{Trellises} & \multicolumn{4}{ |c|}{$\epsilon$} \\ \cline{2-5}
						   & $0.1$ & $0.2$ & $0.3$ &$0.4$ \\ \hline
rsse $[4,1]$ & 1.25dB & 1.3dB & 1.35dB & 1.2dB \\ \hline
rsse $[4,2]$ & $\ll0.1$dB & $0.15$dB & $0.6$dB & $1.1$dB \\ \hline
rsse $[3,3]$ & $1.4$dB & $0.9$dB & $0.6$dB & $0.2$dB \\ \hline
rsse $[4,3]$ & \multicolumn{4}{|c|}{$\ll 0.1$dB} \\ \hline
\end{tabular}
\caption{The SNR loss of different subset trellis configurations to achieve BER$=10^{-4}$ on PR2 channel}
\label{tab_pr2}
\end{table}

\begin{table}
\centering
\begin{tabular}{ |l |l|l|l|l|}
\hline
\multirow{2}{*}{Trellises} & \multicolumn{4}{ |c|}{$\epsilon$} \\ \cline{2-5}
						   & $0.1$ & $0.2$ & $0.3$ &$0.4$ \\ \hline
rsse $[4,3,3]$ & $ 0.1$dB & 0.1dB & 0.05dB & $\ll 0.1$dB \\ \hline
rsse $[4,4,2]$ & $\ll0.1$dB & $0.15$dB & $0.7$dB & $>1$dB  \\ \hline
rsse $[4,3,2]$ & $0.1$dB & $0.25$dB & $0.7$dB & $>1$dB \\ \hline
rsse $[4,2,2]$ & $0.3$dB & \multicolumn{3}{|c|}{$>1$dB} \\ \hline
rsse $[3,3,3]$ & $>1$dB   &   $0.7$dB     &     0.4dB    & $0.05$dB \\ \hline
rsse $[4,3,1]$ & \multicolumn{4}{|c|}{$>1$dB} \\ \hline
\end{tabular}
\caption{The SNR loss of different subset trellis configurations to achieve BER$=10^{-4}$ on EPR4 channel}
\label{tab_epr4}
\end{table}

Table \ref{tab_pr2} and Table \ref{tab_epr4} summarize the performance loss in dB for several subset trellis configurations compared to a ML detector at BER$=10^{-4}$ on channel PR2 and EPR4, respectively. Several conclusions can be drawn from these tables. First, a trellis with fewer states does not necessarily have a worse performance than the one with more states. For example, in Table \ref{tab_epr4} for EPR4 channel, when $\epsilon=0.1$, the configuration $[4,4,2]$ with 32 states outperforms the configuration $[4,3,3]$ with 36 states. Second, the performance of one configuration changes drastically at different ITI levels. One example is the RSSE $[4,4,2]$, which essentially achieves the ML performance at $\epsilon=0.1$, but loses over 1dB for $\epsilon=0.4$. Finally, not all configurations lose more performance at higher ITI. It is interesting to observe that RSSE$[3,3,3]$ with parallel branches can have near optimal performance at $\epsilon=0.4$. Therefore, the pre-selection between parallel branches at every stage is quite realiable. In Section \ref{sec_errorevent} we give an explanation of these observations from the view of error events.  

\subsection{Minimum phase channels}
\begin{figure}
\centering
\subfigure[$\delta=1.3$]{\includegraphics[width=\columnwidth]{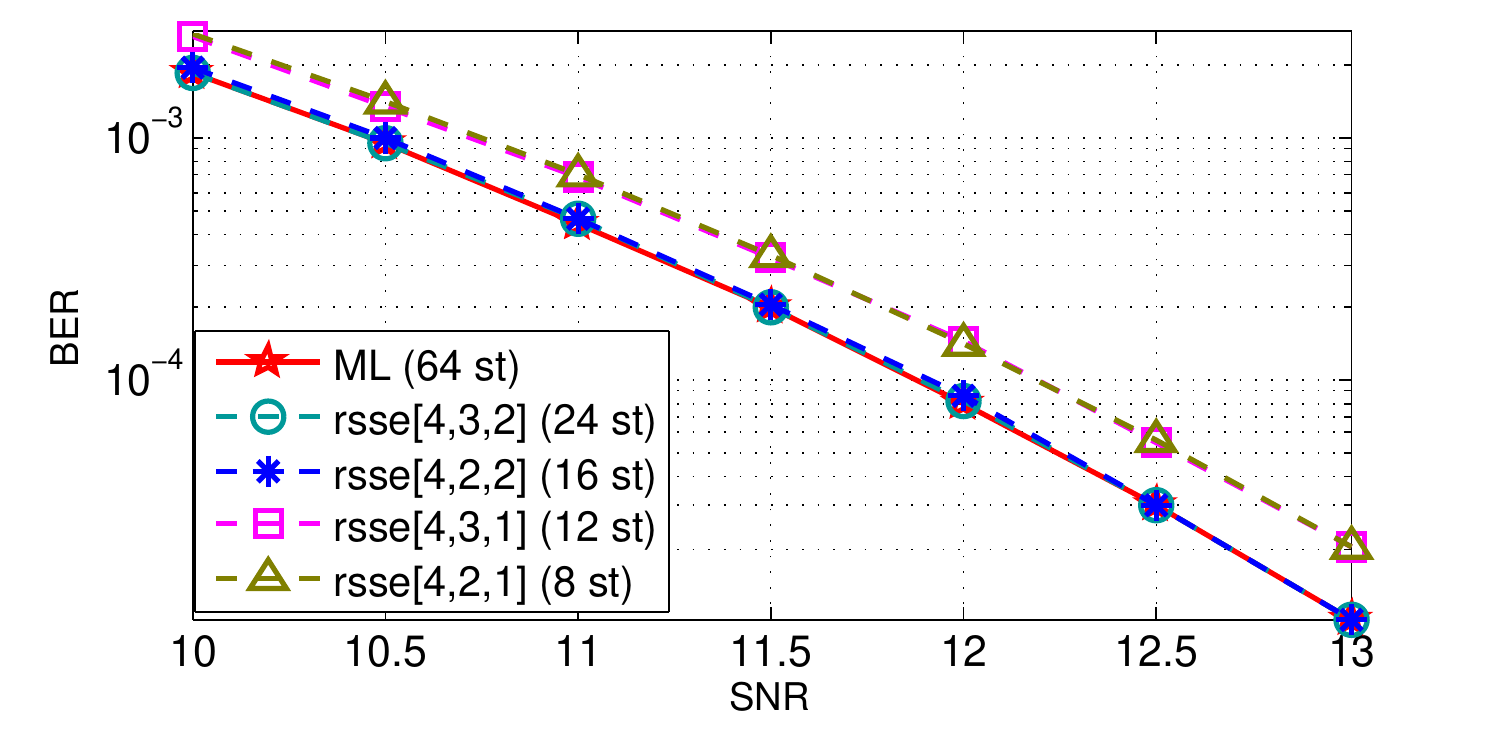}

%\subfigure[Channel I. $h(D)=1+1.6D+1.1D^2+0.4D^3$]{
%\includegraphics[width=\columnwidth]{figures/test}
%\label{sub_type3}
%}
%\subfigure[Channel II. ]{
%\includegraphics[width=\columnwidth]{figures/type5_test}
%\label{sub_type5}
%}
\label{sub_delta3}}

\subfigure[$\delta=1.5$]{
\includegraphics[width=\columnwidth]{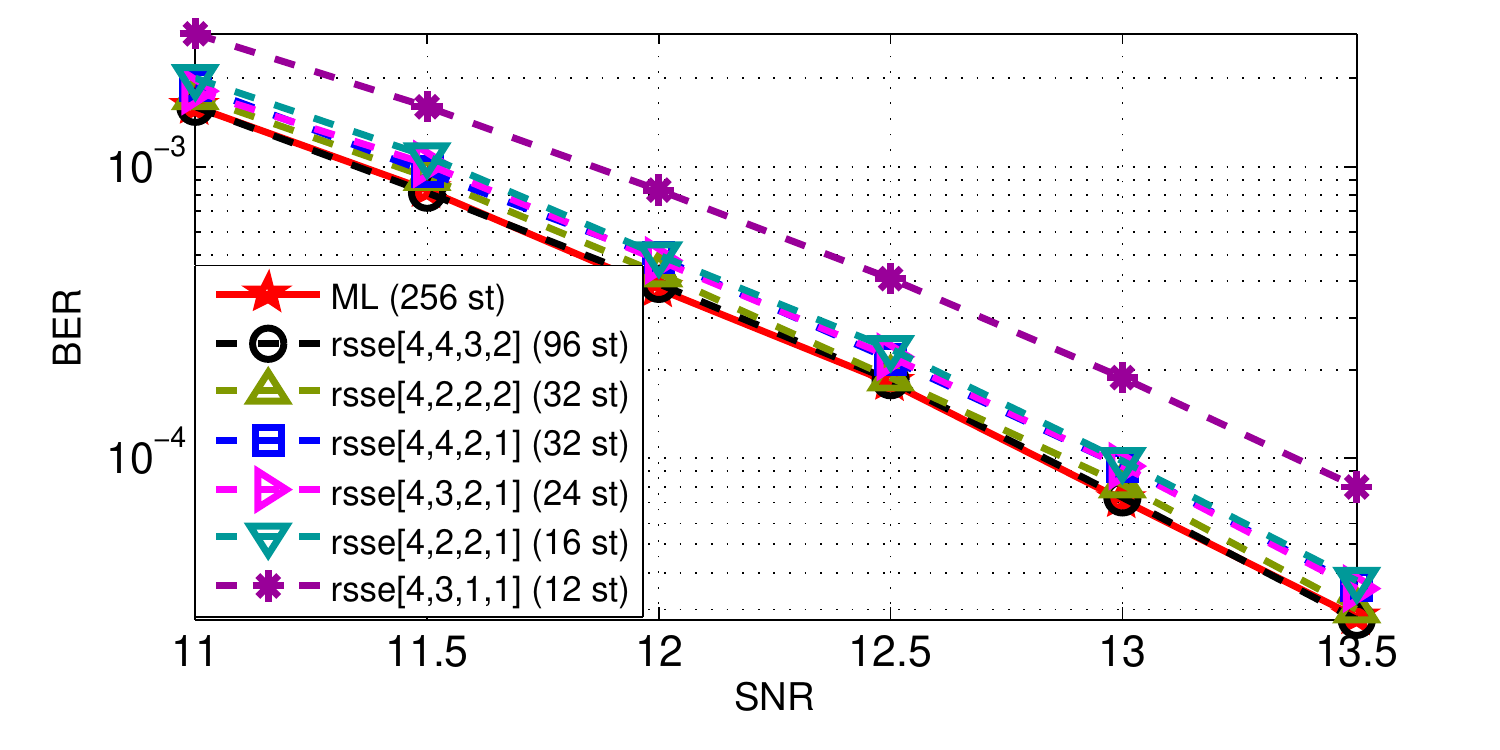}
%\subfigure[Channel I. $h(D)=1+1.6D+1.1D^2+0.4D^3$]{
%\includegraphics[width=\columnwidth]{figures/test}
%\label{sub_type3}
%}
%\subfigure[Channel II. $h(D)=1+1.9D+1.6D^2+0.8D^3+0.3D^4$]{
%\includegraphics[width=\columnwidth]{figures/type5_test}
%\label{sub_type5}
%}

\label{sub_delta5}
}
\caption{Performance comparison between RSSE and ML detector on minimum phase channels at $\epsilon=0.1$. The polynomials are (a) $h(D)=1+1.6D+1.1D^2+0.4D^3$, (b) $h(D)=1+1.9D+1.6D^2+0.8D^3+0.3D^4$.}
\label{fig_miniphase}
\end{figure}

Minimum phase channels can better model the real channel on a disk drive. Assume the transition response of a perpendicular magnetic recording (PMR) disk is $s(t)=V_{\text{max}}\tanh(\frac{2t}{0.579\pi\delta})$, where $V_{\text{max}}$ is the writing voltage and $\delta$ indicates the linear density on one data track. By using the whitened matched filter structure in \cite{Forney72}, we derive two minimum phase channel polynomials: type 1, $h(D)=1+1.6D+1.1D^2+0.4D^3$ for $\delta=1.3$, and type 2, $h(D)=1+1.9D+1.6D^2+0.8D^3+0.3D^4$ for $\delta = 1.5$. These are two commonly used densities in current commercial HDDs. Since the minimum phase condition implies that most of the channel energy is distributed over the most recent samples, the early merge in RSSE can be more reliable compared to the linear phase channel (such as PR2 and EPR4 channel). It is interesting to compare type 1 channel with EPR4, both of which have memory $\nu=3$. As shown in Fig. \ref{sub_delta3}, ML, RSSE$[4,3,2]$, and RSSE$[4,2,2]$ have essentially identical performance. In particular, RSSE$[4,2,2]$ performs much better on the minimum phase channel than on the EPR4 channel (Fig. \ref{sub_epr4_1}), allowing RSSE to achieve near-ML performance with only 16 states. That's only 25\% of the original trellis states. Other more aggressive configurations are also plotted. As can be seen, RSSE with only 8 states can achieve performance that is within 0.3dB of ML detection. 

For the type 2 channel which has higher memory, the RSSE algorithm performs even better as in Fig. \ref{sub_delta5}. It shows that RSSE$[4,2,2,2]$ with 32 states can achieve near-ML performance. That's 12.5\% of the ML trellis states. If 0.1dB loss is permissible, RSSE$[4,2,2,1]$ can be used. So the complexity further drops to 6.25\% of the original ML algorithm.

%\begin{figure}
%\centering
%\subfigure[$h(D)=1+1.6D+1.1D^2+0.4D^3$]{
%\includegraphics[width=0.48\columnwidth]{figures/type3}
%\label{sub_type3}
%}~
%\subfigure[$h(D)=1+1.9D+1.6D^2+0.8D^3+0.3D^4$]{
%\includegraphics[width=0.48\columnwidth]{figures/type5}
%\label{sub_type5}
%}
%\caption{performance on minimum phase channels}
%\label{fig_minphase}
%\end{figure}

%\section{Conclusion}
%\label{sec_conclusion}
%\vspace{-5pt}

\section{Error Event Analysis}
\label{sec_errorevent}
We will use error event analysis to study the performance-complexity tradeoff among different subset trellis configurations. 
 
An error event, $\bfe(D) = \bfz(D)-\hat{\bfz}(D)$, happens if the estimated sequence $\hat{\bfz}(D)$ is different from the transmisted sequence $\bfz(D)$. It is well-known \cite{Forney72} that at high SNR, the error event probability of a trellis-based detector can be approximated by $P_e \approx c\cdot Q(\frac{d_{\text{min}}}{2\sigma})$, where $Q(\cdot)$ is the area under the tail of the standard Gaussian distribution.
\begin{align}
d^2_{\text{min}} = \min_{\bfe(D)} d^2(\bfe(D))
\end{align}
is the minimum distance parameter, and $c$ is a coefficient indicating the average number of error events at distance $d^2_{\text{min}}$. Due to the exponential nature of the $Q$ function, the performance comparison between two detectors can be easily conducted by considering their minimum distance parameter. The error events that lead to $d^2_{\text{min}}$ are called dominant error events.

For the WSSJD detector, due to the unequally scaled noise power, an effective distance measure of a given error event $\bfe(D) = [e^+(D), e^-(D)]$ is defined by
\begin{align}
\small
& d^2_{\text{ML}}(\bfe(D)) \notag \\&=\frac{(1+\epsilon)^2}{2}\|e^+(D)h(D)\|^2+\frac{(1-\epsilon)^2}{2}\|e^-(D)h(D)\|^2.
\label{eq_13}
\end{align}
Comparing equation (\ref{eq_10}) to equation (\ref{eq_13}), we can see that ESPD is proportional to the distance parameter of a single error symbol with $e^+_k = z^+_k-\tilde{z}^+_k$ and $e^-_k = z^-_k-\tilde{z}^-_k$. Recall that the set partition tree is constructed such that $\Delta_1^2, \Delta_2^2$ and $\Delta_3^2$ are the minimum ESPD on each level from the bottom to the top, and $\Delta_1^2, \Delta_2^2$, $\Delta_3^2$ are varying with respect to $\epsilon$. Therefore, the minimum distance parameter of the reduced-state trellis configuration also changes with $\epsilon$, and its trend can be roughly predicted by analyzing the change of minimum ESPD in each $\Omega(k)$. We will give a more detail insights in the following subsections.
The minimum value of $d^2_{\text{ML}}(\bfe(D))$ is abbreviated to $d_{\text{min}}^2$, which is the minimum distance parameter of the 2H2T ML detector. It serves as a benchmark for evaluating the performance loss of RSSE algorithm.

%The effective symbol pair distance is proportional to the distance of length-1 error events. For example, assume that $\bfx_n=(1,1)^T$ is the transmitted symbol and $\tilde{\bfx}_n=(-1,-1)^T$ is the estimated symbol. It corresponds to $\bfz_n=(+2,0)^T$ and $\tilde{\bfz}_n=(-2,0)^T$ in the WSSJD setup, respectively. Let $e_x^a(D)=2D^n$ and $e_x^b(D)=2D^n$ denote the input error sequences.The output error sequences $e_y^a(D)$ and $e_y^b(D)$ are found by
%\begin{align}
%\left[ \begin{array}{c}
%e_y^a(D) \\ e_y^b(D)
%\end{array} \right]
%= \left[ \begin{array}{c c}
%1 & \epsilon \\ \epsilon & 1
%\end{array}\right]\,
%\left[ \begin{array}{c}
%e^a_x(D)h(D) \\ e^b_x(D)h(D)
%\end{array} \right].
%\end{align}
%The distance of this length-1 error event is calculated by
%\begin{align}
%&d^2(e^a_x(D), e^b_x(D)) \notag \\
%& \quad= \|e^a_y(D)\|^2+\|e^b_y(D)\|^2\notag \\
%& \quad=  8(1+\epsilon)^2\|h(D)\|^2 = \Delta_1^2 \|h(D)\|^2 \label{eq_length1} 
%\end{align}
%Therefore the effective symbol pair distance can be viewed as the distance that two input symbols contribute to the output space. 

\subsection{Parallel Branches}
If $J_1<4$, the subset trellis contains parallel branches. So early merge happens at every time instant. Ignoring the error propagation effect, we assume that at time $k$, both the correct and the estimated sequences are at state $\bfs_k$, and $\bfz_{k-i} = \hat{\bfz}_{k-i}$ for all $i = 1, ..., \nu$. At time $k+1$, if $\bfx_{k}, \hat{\bfx}_k \in a_k(1)$, the detector needs to decide a survivor symbol, and discard the other one. If the correct symbol is discarded, this wrong decision can not be reversed in the remaining detection steps. The probability of making a wrong decision in the parallel branch selection is $Q(\frac{h_0d(\bfz_k, \hat{\bfz}_k)}{2\sigma})$, where $d(\bfz_k, \hat{\bfz}_k)$ is the square-root of ESPD between $\bfz_k$ and $\hat{\bfz}_k$. Let $E_1$ be the set of all such length-1 error events due to the parallel branches. Then, $\bfe_k\in E_1$  if and only if there exist two input symbols $ \bfz_{k}, \hat{\bfz}_k \in a_k(1)$ such that $\bfe_k =  \bfz_{k} - \hat{\bfz}_k$. It can be shown that 
 \begin{align}
   & d_{\text{min}}^2(E_1) = \min_{\bfe_k\in E_1}\frac{(1+\epsilon)^2}{2}(h_0e^+_k)^2+\frac{(1-\epsilon)^2}{2}(h_0e^-_k)^2 \notag\\
   & \quad= \left\{ 
    \begin{array}{l l}
      8h_0^2(1+\epsilon)^2 = h_0^2\Delta_1^2 & \quad J_1=3\\
      8h_0^2(1-\epsilon)^2 = h_0^2\Delta_2^2 & \quad J_1 = 2.
    \end{array} \right
    .
    \label{eq_error1}
  \end{align}

The existence of parallel branches will not significantly degrade the performance if it can achieve the same minimum distance as MLSE, i.e., $d_{\text{min}}^2(E_1) \geq d^2_{\text{min}}$. For the dicode channel, 
 \begin{align}
  d_{\text{min}}^2 = \left\{ 
   \begin{array}{l l}
     8(1+\epsilon^2) & \quad \text{if $0\leq \epsilon \leq 2-\sqrt{3}$}\\
     16(1-\epsilon)^2 & \quad \text{if $2-\sqrt3\leq \epsilon \leq 1/2$}.
   \end{array} \right
   .
   \label{eq_mlse}
 \end{align}
Therefore, for the 3-state subset trellis, $d_{\text{min}}^2(E_1) \geq d_{\text{min}}$ for all $\epsilon$, leading to a performance close to the MLSE as shown in Figs. \ref{fig_dicode_result}. Moreover, as $\epsilon$ increases, $d_{\text{min}}^2(E_1)$ becomes much larger than $d_{\text{min}}$, making the effect of these length-1 error events negligible. So we observe that the BER curve of the 3-state RSSE almost overlaps with that of MLSE at $\epsilon=0.3$. In contrast to the 3-state trellis, in the 2-state trellis  $d_{\text{min}}^2(E_1) < d_{\text{min}}$ for all $\epsilon$, resulting in worse performance.

As for channel PR2 and EPR4,  their $d^2_{\text{min}}$ is given by
 \begin{align}
  d_{\text{min}}^2 = \left\{ 
   \begin{array}{l l}
     16(1+\epsilon^2) & \quad \text{if $0\leq \epsilon \leq 2-\sqrt{3}$}\\
     32(1-\epsilon)^2 & \quad \text{if $2-\sqrt3\leq \epsilon \leq 1/2$}.
   \end{array} \right
   .
   \label{eq_pr2_mlse}
 \end{align}
Consider a subset trellis with $J_1=3$. For $\epsilon>\frac{1}{3}$, $d_{\text{min}}^2(E_1)$ is strictly larger than $d^2_{\text{min}}$. Therefore the error events in $E_1$ are not the dominant ones. As shown in Tables \ref{tab_pr2} and \ref{tab_epr4}, RSSE$[3,3]$ and RSSE$[3,3,3]$ perform very close to their relevant MLSE, respectively.
 
\subsection{Early Merging Condition}
We next try to identify longer RSSE error events.
Assume $\bfz(D)$ and $\hat{\bfz}(D)$ are the correct input sequence and the estimated sequence, respectively, and their decoding paths are merged at times $k_1$ and $k_2$ and unmerged in between. Let $E$ denote the set of all error events ending at time $k_2$, where the starting position $k_1$ is arbitrary. According to \cite{rsse88}, an error event $\bfe(D)\in E$ if and only if the following hold.
\begin{enumerate}
\item $\bfe_{k_1} $ is non-zero.
\item The last $\nu$ elements, $[\bfe_{k_2-\nu}, \cdots, \bfe_{k_2-1}]$, should satisfy the ``\textbf{merging condition}", i.e., $\bfe_{k_2-k} = \bfz_{k_2-k}-\hat{\bfz}_{k_2-k}$ where $\bfz_{k_2-k}$ and $\hat{\bfz}_{k_2-k}$ belong to the same subset in the partition $\Omega(k)$ for all $k=1, ..., \nu$.
\item No earlier $\nu$ elements satisfy the merging condition.
\end{enumerate}

In MLSE, the merging condition requires $\bfe_{k_2-k} = 0$ for $k=1, ..., \nu$. However, this is not the case in RSSE. We call the error events $\bfe(D)\in E$ whose last $\nu$ elements are not all zero the ``\textbf{early merged error events}'', denoted $E^r$. Clearly $E_1 \subseteq E^r$. The set of remaining error events, $E^m = E\backslash E^r$, contains the ones that are not affected by the RSSE algorithm. They have the same distance parameters as in MLSE.

How to decide if a given error event belongs to $E^r$? In the following, we give a necessary and sufficient condition for $\bfe(D) \in E^r$, which is called the ``\textbf{early merging condition}". To save space, the error symbols are indexed by digits shown in Table \ref{tab_err_symbol}. Several terminologies are introduced first.
%error alphabet of WSSJD system:
%\begin{table}[!h]
%\centering
%\begin{tabular}{l| l}
%\hline 
%$E_d$ & $(+4,0),(-4,0),(0,+4),(0,-4)$ \\
%\hline 
%$E_s$ & $(+2,+2),(+2,-2),(-2,+2),(-2,-2)$ \\
%\hline
%\end{tabular}
%\end{table}
\begin{table}
	\centering
\small 
\begin{tabular}{ c | l   l }
index & $(e^+_k, e^-_k)$ & $(e^a_k, e^b_k)$ \\ \hline \hline
$0$ & $(0,0)$ & $(0,0)$ \\
$1$ &  $(4, 0)$ &$(2,2)$  \\
$2$ & $(-4,0)$ & $(-2,-2)$ \\
$3$ &  $(0,4)$ & $(2,-2)$ \\
$4$ & $(0,-4)$  & $(-2,2)$ \\
$5$ & $(2,2)$   & $(2,0)$\\
$6$ & $(-2,-2)$ & $(-2,0)$ \\
$7$ & $(2,-2)$  &$(0,2)$\\
$8$ & $(-2,2)$  & $(0,-2)$ \\
\hline
\end{tabular}
\caption{Error symbols by index}
\label{tab_err_symbol}
\end{table}

For a partition $\Omega$ of the input constellation,
the set of \textbf{intrasubset errors}, denoted by $\mathcal{E}_{a}(\Omega)$, is a collection of  error symbols such that if there exist two inputs $\bfz,\hat{\bfz}$ satisfying the condition that $\bfe = \bfz - \hat{\bfz}$ and $\bfz, \hat{\bfz}$ belong to the same subset in $\Omega$, then $\bfe \in \mathcal{E}_{a}(\Omega)$. Similarly, the set of \textbf{intersubset errors}, denoted by $\mathcal{E}_{b}(\Omega)$, is a collection of error symbols such that if there exist two inputs $\bfz,\hat{\bfz}$ satisfying the condition that $\bfe = \bfz - \hat{\bfz}$ and $\bfz, \hat{\bfz}$ belong to two different subsets in $\Omega$, then $\bfe \in \mathcal{E}_{b}(\Omega)$. 
\begin{proposition}
\label{prop_1}
For the proposed set partition tree in Fig. \ref{fig_tree}, $\mathcal{E}_{a}(L_i)\cap \mathcal{E}_{b}(L_i) = \emptyset$ for $i = 1,2,3,4$.
\end{proposition}
\textbf{Proof}. We prove the claim by enumeration.
\begin{enumerate}
\item For $\Omega=L_1$, all error symbols are intrasubset errors since there is only one subset.
\item For $\Omega=L_2$, $\mathcal{E}_{a}(L_2) = \{0,1,2,3,4\}$,  $\mathcal{E}_{b}(L_2) = \{5,6,7,8\}$.
\item For $\Omega = L_3$, $\mathcal{E}_{a}(L_3) = \{0,1,2\}$, $\mathcal{E}_{b}(L_3) = \{3,4,5,6,7,8\}$.
\item For $\Omega=L_4$, all non-zero error symbols are intersubset errors, so $\mathcal{E}_{a}(L_4)=\{0\}$. \hfill $\Box$

\end{enumerate}
\begin{proposition}
\label{prop_2}
(Early merging condition) \\
An error event $\bfe(D)\in E^r$ if and only if the last $\nu$ elements are not all zero symbols, and satisfy $\bfe_{k_2-k}\in \mathcal{E}_a(\Omega(k))$ for all $k=1, ..., \nu$, and no previous $\nu$-tuple satisfies the condition.
\end{proposition}
\textbf{Proof.} Given $\bfe(D)\in E^r$, it is straightforward from the definition of ``merging condition" that the last $\nu$ elements must be intrasubset error symbols in the corresponding patition $\Omega(k)$. On the other hand, if $\bfe_{k_2-k}\in \mathcal{E}_a(\Omega(k))$ for all $k=1, ..., \nu$, by Proposition \ref{prop_1}, the sequences $\bfz(D)$ and $\hat{\bfz}(D)$ that produce $\bfe(D)$ must satisfy that $\bfz_{k_2-k}$ and $\hat{\bfz}_{k_2-k}$ belong to the same subset in $\Omega(k)$. Therefore $\bfz(D)$ and $\hat{\bfz}(D)$ are merged at $k_2$, and $\bfe(D)\in E^r$. \hfill $\Box$
\begin{remark}
Notice that Proposition \ref{prop_1} is also true for the QAM Ungerboeck set partition tree. So Proposition \ref{prop_2} also applies to the original RSSE setup.
\end{remark}
\begin{remark}
The single track error events are not affected by RSSE algorithm if $J_i>1$ for all $i=1, ...,\nu $.
\end{remark}

%Similarly, in RSSE, an error event extends from $k_1$ to $k_2$ if two sequences are merged at $k_1$ and $k_2$ and not merged in between on the subset trellis, i.e. $\bfs_{k_1}=\hat{\bfs}_{k-1}$ and $\bfs_{k_2}=\hat{\bfs}_{k_2}$, but $\bfs_{k}\neq\hat{\bfs}_{k}$ for $k_1<k<k_2$. Notice that having the same subset state at time $k$ only implies that the symbol $\bfz_k-i$ and $\hat{\bfz}_k-i$ belongs to the same subset for $1\leq i \leq \nu$. (example). If at the starting time $k_1$, the path history store a wrong symbol, i.e. the estimated symbol $\hat{\bfz}_{k_1-i}\neq \bfz_{k_1-i}$ for some $i\in[1,\nu]$, the decision feedback scheme process error propagation. Fortunately the effect of error propagation is small\cite{rsse88}. In the following analysis when an error event starts from time $k_1$, we assume there is no error propagation, i.i. $\bfe_{k_1-1}=\bfe_{k_1-2}=\dots=\bfe_{k_1-\nu}=0$.
Assume $\bfe(D)\in E^r$ and starts from $k_1$. The distance parameter of $\bfe(D)$ is given by
\begin{align}
\small
& d_{r}^2(\bfe(D)) \notag \\
& =\frac{(1+\epsilon)^2}{2}\sum\limits_{k=k_1}^{k_2} (\sum\limits_{i=0}^{\nu} e^+_{k-i}h_i)^2 + \frac{(1-\epsilon)^2}{2}\sum\limits_{k=k_1}^{k_2} (\sum\limits_{i=0}^{\nu} e^-_{k-i}h_i)^2
\label{eq_17}
\end{align}
The distance parameter measured by equation (\ref{eq_17}) is always smaller than or equal to that measured by equation (\ref{eq_13}) \cite{rsse88}. The decrement is the price we paid for using the reduced-state trellis. An example is given to illustrate the difference.
\begin{example}
\label{exam_1}
Consider the PR2 channel, and assume $\epsilon=0.1$. Assume a single error $\bfe_{k_1} = (4,0)$ happens at time $k_1$. In MLSE, the distance parameter contributed by $\bfe_{k_1}$ is $\frac{(1+0.1)^2}{2}\cdot (4^2 + 8^2 + 4^2) = 58.08$. However, if RSSE$[4,3]$ is used, this error event will be early merged at time $k_1+1$, since $(4,0)\in \mathcal{E}_a(\Omega(2))$. Therefore the distance parameter of $(4,0)$ in RSSE$[4,3]$ is reduced to $\frac{(1+0.1)^2}{2}\cdot (4^2 + 8^2 ) = 48.4$.
\end{example}

\subsection{Error state diagram}
\begin{table*}
\centering
\caption{The dominant RSSE error events for channel $[1, 1.6, 1.1, 0.4]$}
\label{tab_min_phase}
\begin{tabular}{|l|l|l|l|l|}
\hline
		 & $\epsilon=0.1$ &$\epsilon=0.2$ & $\epsilon=0.3$ & $\epsilon=0.4$ \\ \cline{2-5}
		& $d_{\text{min}}^2 = 9.1304$ & $d_{\text{min}}^2 = 9.4016$ & $d_{\text{min}}^2 = 8.8592$ & $d_{\text{min}}^2 = 6.5088$ \\ \hline \hline
RSSE$[3,3,3]$ (27 st) & $[1]/9.6800$ & $[5\ 2\ 1\ 2]/ 10.7168$ & $[5\ 2\ 1\ 2]/ 10.0028$ & $[5\ 2\ 1\ 2]/ 9.5472$ \\ \hline
RSSE$[3,3,2]$ (18 st) & $[1]/9.6800$ & $[5\ 2\ 1\ 2]/ 10.7168$ & $[3\ 4\ 0\ 0]/8.2320$ & $[3\ 4\ 0\ 0]/6.0480$ \\ \hline
RSSE$[4,3,2]$ (24 st)& $[3\ 4\ 0\ 0]/13.6080$	  & $[3\ 4\ 0\ 0]/10.7520$ & $[3\ 4\ 0\ 0]/8.2320$ & $[3\ 4\ 0\ 0]/6.0480$	\\ \hline
RSSE$[4,2,2]$ (16 st) & $[3 \ 4\ 0]/ 10.4328$ &  $[3 \ 4\ 0]/ 8.2432$ &  $[3 \ 4\ 0]/ 6.3112$ & $[3 \ 4\ 0]/ 4.6368$ \\ \hline
RSSE$[3,2,2]$ (12 st) & $[1]/9.6800$ &  $[3 \ 4\ 0]/ 8.2432$ &  $[3 \ 4\ 0]/ 6.3112$ & $[3 \ 4\ 0]/ 4.6368$ \\ \hline
RSSE$[4,2,1]$ (8 st) &  &  $[3 \ 4\ 0]/ 8.2432$ &  $[3 \ 4\ 0]/ 6.3112$ & $[3 \ 4\ 0]/ 4.6368$ \\
&$[5 \ 6\ 0 \ 0]/ 8.4840$ & $[5\ 6\ 0\ 0]/8.7360$ & & \\ \hline
\end{tabular}

\end{table*}

\begin{table*}
\centering
\caption{The dominant RSSE error events for PR2 channel }
\label{tab_pr2_err}
\begin{tabular}{|l|l|l|l|l|}
\hline
		 & $\epsilon=0.1$ &$\epsilon=0.2$ & $\epsilon=0.3$ & $\epsilon=0.4$ \\ \cline{2-5}
		& $d_{\text{min}}^2 = 16.16$ & $d_{\text{min}}^2 = 16.64$ & $d_{\text{min}}^2 = 15.68$ & $d_{\text{min}}^2 =11.52$ \\ \hline \hline
RSSE$[4,3]$ (12 st)  & $[5\ (2\ 1)^\infty \ 0]/ 24.24 $ & $[5\ (2\ 1)^\infty \ 0]/ 24.96 $ &$[5\ (2\ 1)^\infty \ 0]/ 26.16 $ & $[5\ (2\ 1)^\infty \ 0]/ 27.84 $ \\ 
                         & $[5\ (2\ 1)^\infty \ 2 \ 0]/ 24.24 $ & $[5\ (2\ 1)^\infty\ 2 \ 0]/ 24.96 $ &$[5\ (2\ 1)^\infty \ 2\ 0]/ 26.16 $ & $[5\ (2\ 1)^\infty\ 2 \ 0]/ 27.84 $ \\ \hline
RSSE$[4,2]$ (8 st) & $[(3\ 4)^\infty\ 0]/19.44$	  & $[(3\ 4)^\infty\ 0]/15.36$ & $[(3\ 4)^\infty\ 0]/11.76$ & $[(3\ 4)^\infty\ 0]/8.64$	\\ 
                        & $[(3\ 4)^\infty\ 3\ 0]/19.44$	  & $[(3\ 4)^\infty\ 3\ 0]/15.36$ & $[(3\ 4)^\infty\ 3\ 0]/11.76$ & $[(3\ 4)^\infty\ 3\ 0]/8.64$	\\ \hline
RSSE$[3,3]$ (9 st) & $[1]/9.68$ &  $[1]/ 11.52$ &  $[1]/ 13.52$&  \\
						& $[5\ 2\ 1]/14.56$ &  $[5\ 2\ 1]/13.44$ &  $[5\ 2\ 1]/12.64$ & $[5\ 2\ 1]/12.16$  \\
						& $[(5\ 6)^\infty \ 0\ 2]/16.16$ &  $[(5\ 6)^\infty \ 0\ 2]/16.64$ &   &   \\
						& $[(5\ 6)^\infty \ 5\ 0\ 1]/16.16$ &  $[(5\ 6)^\infty \ 5\ 0\ 1]/16.64$ &   &   \\
						& $[(5\ 6)^\infty \ 2\ 1]/16.16$ &  $[(5\ 6)^\infty \ 2\ 1]/16.64$ &   &   \\
						& $[(5\ 6)^\infty \ 5\ 1\ 2]/16.16$ &  $[(5\ 6)^\infty \ 5\ 1\ 2]/16.64$ &   &   \\ \hline
\end{tabular}
\end{table*}

\begin{table*}
%\centering
\caption{The dominant RSSE error events for EPR4 channel }
\label{tab_epr4_err}
\hspace{-5mm}
\begin{tabular}{|l|l|l|l|l|}
\hline
		 & $\epsilon=0.1$ &$\epsilon=0.2$ & $\epsilon=0.3$ & $\epsilon=0.4$ \\ \cline{2-5}
		& $d_{\text{min}}^2 = 16.16$ & $d_{\text{min}}^2 = 16.64$ & $d_{\text{min}}^2 = 15.68$ & $d_{\text{min}}^2 =11.52$ \\ \hline \hline
RSSE$[4,3, 3]$ (36 st)  & $[(5\ 0)^\infty \ 1\ 0]/16.16 $ &$[(5\ 0)^\infty \ 1\ 0]/16.64 $&$[(5\ 0)^\infty \ 1\ 0]/17.44 $ & $[(5\ 0)^\infty \ 1\ 0]/18.56 $ \\ \hline
RSSE$[4,3, 2]$ (24 st) & & $[(3\ 0)^\infty\ 0]/15.36$ & $[(3\ 0)^\infty\ 0]/11.76$ &$[(3\ 0)^\infty\ 0]/8.64$ \\ 
 & & $[(3\ 4)^\infty\ 3\ 0\ 0]/15.36$ & $[(3\ 4)^\infty\ 3\ 0\ 0]/11.76$ &$[(3\ 4)^\infty\ 3\ 0\ 0]/8.64$ \\ 
 & & $[3\ 4\ (3\ 4)^\infty\ 0\ 0]/15.36$ & $[3\ 4\ (3\ 4)^\infty\ 0\ 0]/11.76$ &$[3\ 4\ (3\ 4)^\infty\ 0\ 0]/8.64$ \\ 
  & $[(5\ 0)^\infty \ 1\ 0]/16.16 $  & $[(5\ 0)^\infty \ 1\ 0]/16.64 $  &    &  \\ \hline
 RSSE$[3,3, 3]$ (27 st) & $[1]/9.68$ & $[1]/11.52$ &$[1]/13.52$ &  \\  
 & $[5\ 6\ 1\ 2\ 1]/14.56$ &$[5\ 6\ 1\ 2\ 1]/13.44$ & $[5\ 6\ 1\ 2\ 1]/12.64$ &  $[5\ 6\ 1\ 2\ 1]/12.16$ \\  
 & $[5\ 2\ 1\ 2]/16.16$ &$[5\ 2\ 1\ 2]/16.64$ &  &   \\  
  & $[(5\ 0)^\infty\ 0\ 1]/16.16$ &$[(5\ 0)^\infty\ 0\ 1]/16.64$ &  &   \\  
  & $[(5\ 0)^\infty\ 1\ 0]/16.16$ &$[(5\ 0)^\infty\ 1\ 0]/16.64$ &  &   \\  
& $[(5\ 0)^\infty\ 1\ 2]/16.16$ &$[(5\ 0)^\infty\ 1\ 2]/16.64$ &  &   \\  
  & $[5\ 6\ (5\ 6)^\infty\ 0\ 0\ 2]/16.16$ &$[5\ 6\ (5\ 6)^\infty\ 0\ 0\ 2]/16.64$ &  &   \\  
  & $[5\ (6\ 5)^\infty\ 0\ 0\ 1]/16.16$ &$[5\ (6\ 5)^\infty\ 0\ 0\ 1]/16.64$ &  &   \\ 
  & $[5\ 6\ (5\ 6)^\infty\ 1\ 2\ 1]/16.16$ &$[5\ 6\ (5\ 6)^\infty\ 1\ 2\ 1]/16.64$ &  &   \\ 
  & $[5\ (6\ 5)^\infty\ 2\ 1\ 2]/16.16$ &$[5\ (6\ 5)^\infty\ 2\ 1\ 2]/16.64$ &  &   \\ \hline
\end{tabular}
\end{table*}

An error state diagram can be employed to search for the minimum distance and enumerate the dominant error events. Consider a labeled directed graph $G=[V, E]$. The vertex set $V$ is the collection of all possible error states $[\bfe_{k-1},\cdots, \bfe_{k-\nu} ]$, so $|V| = 9^{\nu}$. A state that satisfies the merging condition is called a merging state. For MLSE, the all-zero state is the only merging state. For RSSE, the additional merging states are those which satisfy the early merging condition, called ``early merging states". If $\mathcal{T}$ denote the set of merging states, then $|\mathcal{T}| = \prod_{k=1}^{\nu}|\mathcal{E}_a(\Omega(k))|$, which depends on the trellis configuration. An edge $e\in E$ starts from initial state $i(e) = [\bfe_{k-1},\cdots, \bfe_{k-\nu} ]$ and ends in terminal state $t(e) = [\bfe_{k},\cdots, \bfe_{k-\nu+1} ]$, with input/output label $\bfe_k/ \cal{L}_{\text{out}} $. Here
\begin{align*}
 {\cal{L}}_{\text{out}}=\frac{(1+\epsilon)^2}{2}( \sum\limits_{i=0}^{\nu} e^+_{k-i}h_i)^2 +\frac{(1-\epsilon)^2}{2}(\sum\limits_{i=0}^{\nu}e^-_{k-i}h_i)^2.
\end{align*}
 Notice that all the merging states except the all-zero state are sink nodes, which have no outgoing edges. A path starting from the all-zero state and terminating at the merging state defines an closed error event, and the sum of the output labels of all edges in the path gives the distance parameter of this error event. A closed error event that ends at a non-zero merging state is an early merged error event. The depth first algorithm (Algorithm \ref{ag_1}) can be used to find all the error events that lead to a distance parameter smaller than a given threshold.

In Table \ref{tab_min_phase} we summarize the dominant error events for several RSSE configurations for the type-1 minimum phase channel. We simplify the presentation as follows: if $d^2_\text{min}(E^r)\geq d^2_{\text{min}}$, we only list the early merged error events that lead to $d^2_\text{min}(E^r)$; if $d^2_\text{min}(E^r)<d^2_{\text{min}}$, we list all the early merged error events whose distance parameters are smaller than or equal to $d^2_{\text{min}}$. The table is also simplified by considering the symmetry of the error events, i.e., $\pm(e^a(D), e^b(D))$ and $\pm(e^b(D), e^a(D))$ will produce the same distance parameter, so we group them together and only list the one whose first error symbol has a positive $e^a$ component. As shown in Table \ref{tab_min_phase}, the early merged error events in RSSE$[3,3,3]$ always have greater distance parameter than $d_{\text{min}}^2$, under all ITI levels. Particularly, when $\epsilon=0.1$, $E_1$ produced by parallel branches are the dominant error events. As $\epsilon$ increases, $d^2_{\text{min}}(E_1)$, which is proportional to $\Delta_1^2$, also increases, and $[5,2,1,2]$ becomes the dominant one. For RSSE$[4,3,2]$, the error event $[3,4,0,0]$ is the dominant, and its distance parameter decreases as $\epsilon$ increases. In particular, for $\epsilon=0.3$ and $0.4$, its distance parameter is strictly less than $d_{\text{min}}^2$, so it can be predicted that RSSE$[4,3,2]$ loses performance compared to MLSE under high ITI level. One way to avoid this performance loss is to use RSSE$[4,3,3]$ which prevents the error event $[3,4,0,0]$ from being early merged. RSSE$[4,2,2]$ has near-optimal performance at $\epsilon=0.1$ and starts to have significant performance loss when $\epsilon\geq 0.2$. A more aggressive configuration, RSSE$[4,2,1]$, cannot guarantee near-optimal performance since $d_{\text{min}}^2(E^r)$ is always smaller than $d_{\text{min}}^2$. Therefore, to retain near optimal performance as well as reduce complexity, we may use RSSE$[3,2,2]$ at $\epsilon=0.1$, RSSE$[3,3,2]$ at $\epsilon=0.2$, RSSE$[3,3,3]$ at $\epsilon=0.3$ and $0.4$.

For the PR2 and EPR4 channel, the error state diagrams contain zero cycles, which lead to infinite recursive loops in the error event search. Recall that a zero cycle is a path that starts and ends at the same state, and accumulates zero path metric. By definition, zero cycles do not contain merging states, therefore, the number of zero cycles depends on which reduced-state trellis configuration is used. In Examples \ref{exam_2} and \ref{exam_3}, we summarize the zero cycles for PR2 and EPR4 channels. We follow the notation in \cite{Altekar98} and let $(\bfe_1, \cdots, \bfe_k)^\infty$ represent an infinite periodic sequence with repeated pattern $\bfe_1, \cdots, \bfe_k$. Notice that a periodic sequence of the shifted pattern $(\bfe_i, \cdots, \bfe_k, \bfe_{1}, \cdots, \bfe_{i-1})^\infty$ is equivalent to $(\bfe_1, \cdots, \bfe_k)^\infty$. 
\begin{example}
\label{exam_2}
The zero cycles of MLSE on PR2 channel are $0^\infty$, $(1,2)^\infty$, $(3,4)^\infty$, $(5,6)^\infty$, $(7,8)^\infty$. For RSSE$[3,3]$, since $[2,1]$ and $[1,2]$ are both early merging states, $(1,2)^\infty$ is not a zero cycle, while other zero cycles still exist.
\end{example}
\begin{example}
\label{exam_3}
The zero cycles of MLSE on EPR4 are $(0)^\infty$,$\pm (0,1)^\infty$, $\pm (0,3)$, $\pm (0,5)$, $\pm (0,7)$,  $ \pm (1)^\infty$, $\pm (1,2)^\infty$, $\pm (1,3)^\infty$, $\pm (1,4)^\infty$, $\pm (1,5)^\infty$, $\pm (1,6)^\infty$, $\pm (1,7)^\infty$,  $\pm (1,8)^\infty$, $\pm (3)^\infty$, $\pm (3,4)^\infty$, $\pm (3,5)^\infty$, $\pm (3,6)^\infty$, $\pm (3,7)^\infty$, $\pm (3,8)^\infty$, $\pm (5)^\infty$, $\pm (5, 6)^\infty$, $\pm (5, 7)^\infty$, $\pm (5, 8)^\infty$, $\pm (7)^\infty$, $\pm (7, 8)^\infty$. Here $-(\cdot)^\infty$ represents taking the additive inverse of all symbols inside $(\cdot)$.
\end{example}
\begin{remark}
The zero cycles do not intersect, which means, each state can only be visited by at most one zero cycle. We use $\gamma(s)$ to denote the zero cycle which stars and ends at state $s$, and $\gamma(s,j)$ to be the fragment of the zero cycle from state $s$ to $j$. With an abuse of notation, we also use $\gamma(s,j)$ to represent the sequence of input labels on the fragment. The meaning will be clear according to the context.
\end{remark}

Let $\mathcal{Z}$ denote the set of all states that are visited by some zero-cycles, and let $\mathcal{T}$ be the set of all merging states. A 2-step algorithm introduced in \cite{Altekar98} could be used to search for the dominant error events, with a few modification by considering the additional early merging states in RSSE. The procedure is summarized below.
\begin{enumerate}
\item Given a threshold $d^2_{\text{min}}$, apply a depth-first search algorithm to search for all the error fragments $\underline{\bfe}$ with $d^2(\underline{\bfe})\leq d^2_{\text{max}}$  that start from a state in $\cal{Z}$ and end up at a state in $\mathcal{Z}\cup \mathcal{T}$ without having visited $\cal{Z}$ in between. The distance parameter of an error fragment $d^2(\underline{\bfe})$ is the sum of the output labels.
\item Construct a new graph $F$ whose vertices are the states in $\mathcal{Z}\cup \mathcal{T}$.
The edges in $F$ are found as follows. If there is an error fragment $\underline{\bfe}$ starting from state $i$ and ending up at state $j$, then for each state $j'\in \gamma(j)$, there is an edge from state $i$ to $j'$. The input label of the edge is $\underline{\bfe}+\gamma(j,j')$, and the output label is $d^2(\underline{\bfe})$, since the path metric from $j$ to $j'$ is zero. Parallel edges are allowed. 
\item A depth-first search on $F$, which is similar to algorithm \ref{ag_1}, can be used to search and list all the closed error events whose distance parameters are less than $d_{\text{max}}^2$.
\end{enumerate}
Tables \ref{tab_pr2_err} and \ref{tab_epr4_err} list the dominant RSSE error events for some different trellis configurations on PR2 and EPR4 channels, respectively. They are constructed in the same manner as Table \ref{tab_min_phase}. The tables show a good match with the simulation results in Table \ref{tab_pr2} and \ref{tab_epr4}. 

\begin{algorithm}[!t]
\caption{search for closed error events}
\begin{algorithmic}[1]
\State \textbf{function} search$(\text{state }s,\text{path } \underline{p}, d^2(\underline{p})) $
\If{$d^2(\underline{p})<d^2_{\text{max}}$}
\If{$s \in \mathcal{T}$ }
\If{$d^2(\underline{p})>0$}
\State print $\underline{p}$
\EndIf
\Else 
\For{each edge $e$ leaving $s$}
\State $\underline{p}' = \underline{p}+\mathcal{L}_{\text{in}}(e)$
\State search$(t(e), \underline{p}', d^2(\underline{p})+\mathcal{L}_{\text{out}}(e))$
\EndFor
\EndIf
\EndIf
\end{algorithmic}
\label{ag_1}
\end{algorithm}

\section{Asymmetric 2H2T System}
\label{sec_asymmetric}
In an asymmetric 2H2T system, the ITI levels sensed by two heads are different, i.e.,
\begin{align}
\left[ \begin{array}{c}
r^a(D) \\ r^b(D)
\end{array} \right]
= & \left[ \begin{array}{c c}
1 & \epsilon-\Delta\epsilon \\ \epsilon+\Delta\epsilon & 1
\end{array} \right]
\left[ \begin{array}{c}
x^a(D)h(D) \\ x^b(D)h(D)
\end{array} \right]\notag\\ 
&+ \left[ \begin{array}{c}
n^a(D) \\ n^b(D)
\end{array} \right].
\label{eq_asy}
\end{align}
Without loss of generality, we assume $0<\Delta\epsilon<\epsilon$.
\begin{proposition}
The minimum distance parameter of the ML detector on the asymmetric 2H2T channel is
 \begin{align}
  d_{\text{min}}^2 = \left\{ 
   \begin{array}{l l}
     (1+(\epsilon-\Delta\epsilon)^2)d_0^2 & \quad \text{if $ (\epsilon+\Delta\epsilon)^2-4\epsilon+1 \geq 0$}\\
     2[(1-\epsilon)^2+\Delta\epsilon^2] d_0^2 & \quad \text{otherwise}
   \end{array} \right.
   \label{eq_3}
 \end{align}
\end{proposition}
\textbf{Proof.}
Assume $e^a(D)$ and $e^b(D)$ are the error events on track $a$ and $b$, respectively. Let
\begin{align}
\left[ \begin{array}{c}
e_y^a(D) \\ e_y^b(D)
\end{array} \right]
= \left[ \begin{array}{c c}
1 & \epsilon-\Delta\epsilon \\ \epsilon+\Delta\epsilon & 1
\end{array} \right]
\left[ \begin{array}{c}
e^a(D)h(D) \\ e^b(D)h(D)
\end{array} \right].
\end{align}
For the asymmetric channel model, the distance parameter of error event $(e^a(D), e^b(D))$ is calculated by
\begin{align}
d^2(e^a(D), e^b(D)) &= \|e_y^a(D)\|^2 +  \|e_y^b(D)\|^2\notag \\
&= [1+(\epsilon+\Delta\epsilon)^2] \|e^a(D)h(D)\|^2 \notag\\ 
&\quad+  [1+(\epsilon-\Delta\epsilon)^2] \|e^b(D)h(D)\|^2  \notag\\
& \quad+4\epsilon \left< e^a(D)h(D), e^b(D)h(D) \right>.
 \label{eq_21}
\end{align}
We find the minimum value of equation (\ref{eq_21}) by discussing single track error events and double track error events separately.\\ 
\underline{\textit{Single track error event}}. Assume $e^b(D) = 0$,  
\begin{align}
d^2(e^a(D), 0) &= [1+(\epsilon+\Delta\epsilon)^2]\|e^a(D)h(D)\|^2 \\ & \geq [1+(\epsilon+\Delta\epsilon)^2]d_0^2
\end{align}
Similarly, in the case that $e^a(D) = 0$,  
\begin{align}
d^2(0, e^b(D)) \geq [1+(\epsilon-\Delta\epsilon)^2]d_0^2
\end{align}
Therefore, the minimum distance parameter associated with single track error events is $[1+(\epsilon-\Delta\epsilon)^2]d_0^2$, obtained when $e^a(D)=0$ and $e^b(D)$ achieves $d^2_0$ on $h(D)$.\\
\underline{\textit{Double track error event}}.
Consider equation (\ref{eq_21}). By the Cauchy-Schwartz inequality,
\begin{align}
&\left<e^a(D)h(D), e^b(D)h(D)\right> \notag\\&\quad\geq -\|e^a(D)h(D)\|\cdot\|e^b(D)h(D)\| \notag\\
&\quad\geq -\frac{1}{2}\left(\|e^a(D)h(D)\|^2+\|e^b(D)h(D)\|^2\right),
\end{align}
equation (\ref{eq_21}) becomes
\begin{align}
d^2(e^a(D), e^b(D)) &\geq [1+(\epsilon+\Delta\epsilon)^2-2\epsilon] \|e^a(D)h(D)\|^2 \notag\\ &\quad+  [1+(\epsilon-\Delta\epsilon)^2-2\epsilon] \|e^b(D)h(D)\|^2  \notag\\
&\geq 2[(1-\epsilon)^2+\Delta\epsilon^2] d_0^2.
\end{align}
The lower bound can be achieved by the error events which satisfy $e^a(D)=-e^b(D)$, where $e^a(D)$ achieves $d_0^2$ on $h(D)$.
 
\hfill $\Box$

Relative to a symmetric channel, on the asymmetric channel the minimum distance parameter of single track error events is always decreased, while that of double track error events is always increased.

\vspace{1em}
To be consistent with the previous discussions, we analyze the performance of RSSE on the asymmetric channel by considering the transformed system. After the coordinate transformation in both input and output spaces, the asymmetric 2H2T channel becomes 
\begin{align}
\left[\! \begin{array}{c}
r^+(D) \\ r^-(D)
\end{array} \!\right]
= \left[ \!\begin{array}{c c}
1 & \frac{\Delta\epsilon}{1+\epsilon} \\ \frac{\Delta\epsilon}{\epsilon-1} & 1
\end{array}\! \right]
\left[ \!\begin{array}{c}
z^+(D)h(D) \\ z^-(D)h(D)
\end{array}\! \right] + 
\left[ \!\begin{array}{c}
n^+(D) \\ n^-(D)
\end{array} \!\right],
\label{eq_trans_asy}
\end{align}
where the transformed channel inputs, outputs and noises are obtained from equation (\ref{eq_z}), (\ref{eq_r}), and (\ref{eq_n}), respectively.

\begin{table}
\begin{tabular}{|l|c|c|c|}
\hline 
 & $\Delta\epsilon=0$ & $\Delta\epsilon=0.05$ & $\Delta\epsilon=0.1$ \\ \cline{2-4}
 & $d^2_{\text{min}} = 16.16$ & $d^2_{\text{min}} = 16.04$ & $d^2_{\text{min}} = 16$ \\ \hline
 RSSE$[4,4,3]$ & 22.64 & 22.70 & 22.88\\ \hline
 RSSE$[4,4,2]$ & 19.44 & 19.50 & 19.68\\ \hline
 RSSE$[4,4,1]$ & 12.12 & 12.03 & 12.00\\ \hline
 RSSE$[4,3,3]$ & 16.16 & 16.20 & 16.32 \\ \hline
 RSSE$[4,3,2]$ & 16.16 & 16.20 & 16.32 \\ \hline
 RSSE$[3,3,3]$ & 9.68 & 9.70 & 9.76 \\ \hline
\end{tabular}
\caption{$d^2_{\text{min}}(E^r)$ for asymmetric 2H2T EPR4 channel with $\epsilon=0.1$ and various $\Delta \epsilon$.}
\label{tab_asy1}
\end{table}

\begin{table}
\begin{tabular}{|l|c|c|c|}
\hline 
 & $\Delta\epsilon=0$ & $\Delta\epsilon=0.05$ & $\Delta\epsilon=0.1$ \\ \cline{2-4}
 & \small $d^2_{\text{min}} = 11.52$ & \small$d^2_{\text{min}} = 11.60$ & \small$d^2_{\text{min}} = 11.84$ \\ \hline
 RSSE$[4,4,3]$ & 21.44 & 21.50 & 21.68\\ \hline
 RSSE$[4,4,2]$ & 8.64 & 8.7 & 8.88\\ \hline
 RSSE$[4,4,1]$ & 8.64 & 8.7 & 8.88\\ \hline
 RSSE$[4,3,3]$ & 18.56 & 18.60 & 18.72 \\ \hline
 RSSE$[4,3,2]$ & 8.64 & 8.7 & 8.88 \\ \hline
 RSSE$[3,3,3]$ & 12.16 & 12.20 & 12.32 \\ \hline
\end{tabular}
\caption{$d^2_{\text{min}}(E^r)$ for asymmetric 2H2T EPR4 channel with $\epsilon=0.4$ and various $\Delta \epsilon$.}
\label{tab_asy2}
\end{table}

The WSSJD on the asymmetric 2H2T channel follows the same procedure as in Section \ref{subsec_wssjd}. The only difference is that in the asymmetric system, the noiseless channel outputs become
\begin{align}
y^+_n=\sum_{i=0}^{\nu}h_iz^+_{n-i}+ \frac{\Delta\epsilon}{1+\epsilon}\sum_{i=0}^{\nu}h_iz^-_{n-i}
\end{align}
and
\begin{align}
y^-_n=\sum_{i=0}^{\nu}h_iz^-_{n-i}+ \frac{\Delta\epsilon}{\epsilon-1}\sum_{i=0}^{\nu}h_iz^+_{n-i}.
\end{align}

When applying RSSE to the asymmetric channel, we use the same set partition tree as in Fig. \ref{fig_tree}, and investigate its performance trend by means of both simulation and error event analysis. 
We first consider the change of trellis with parallel branches. Notice that the thresholding in parallel branch selection does not work for the asymmetric channel, so branch metrics should be calculated explicitly to compare. Assuming $J_1>1$, the effective squared distance between two parallel branches coming from the same trellis state is
\begin{align}
d^2(\bfe_k\in E_1) & = \frac{(1+\epsilon)^2}{2}(h_0e^+_k + \frac{\Delta\epsilon}{1+\epsilon}h_0e^-_k)^2 \notag \\
&\quad +  \frac{(1-\epsilon)^2}{2}(\frac{\Delta\epsilon}{\epsilon-1}h_0e^+_k + h_0e^-_k)^2 \label{eq_30}\\
& = \frac{h_0^2(1+\epsilon)^2}{2}[(e^+_k)^2 + \frac{\Delta\epsilon^2}{(1+\epsilon)^2}(e^-_k)^2] \notag \\
&\quad +  \frac{h_0^2(1-\epsilon)^2}{2}[\frac{\Delta\epsilon^2}{(\epsilon-1)^2}(e^+_k)^2 + (e^-_k)^2]. \label{eq_31}
\end{align}
The second equality follows from the fact that when $J_1=2$ or $J_1=3$, $\bfe_k$ always has a zero component, so $e^+_ke^-_k=0$. The minimum distance parameter associated with $E_1$ is
\begin{align}
    d_{\text{min,asy}}^2(E_1) &= \min_{\bfe_k\in E_1} d^2(\bfe_k\in E_1)\notag\\
   & = \left\{ 
    \begin{array}{l l}
      h_0^2\Delta_1^2+8\Delta\epsilon^2h_0^2 & \quad J_1=3\\
      h_0^2\Delta_2^2+8\Delta\epsilon^2h_0^2 & \quad J_1 =2.
    \end{array} \right.
    \label{eq_32}
\end{align}
Compared to the symmetric case, $d^2_{\text{min,asy}}(E_1)$ is increased both for $J_1=2$ and $J_1=3$. For the asymmetric 2H2T system with dicode channel target, the minimum distance parameter of MLSE is given in equation (\ref{eq_3}) by letting $d_0^2 = 8$. It is easy to verify that when $J_1=3$, $d^2_{\text{min,asy}}(E_1)>d^2_{\text{min}}$ for all non-zero $\epsilon$ and $\Delta \epsilon$.

Let $e^+(D)=e^a(D)+e^b(D)$ and $e^-(D)=e^a(D)-e^b(D)$ be the error events of the transformed system. The squared distance given in equation (\ref{eq_21}) can be expressed as
\begin{align}
& d^2(e^+(D), e^-(D)) \notag\\ & \quad = \frac{(1+\epsilon)^2}{2}\|e^+(D)h(D) + \frac{\Delta\epsilon}{1+\epsilon}e^-(D)h(D)\|^2 \notag \\
&\quad +  \frac{(1-\epsilon)^2}{2}\|\frac{\Delta\epsilon}{\epsilon-1}e^+(D)h(D) + e^-(D)h(D)\|^2. \label{eq_34}
\end{align}
Then the error state diagram and the error event search algorithm introduced in Section \ref{sec_errorevent} can be applied to the asymmetric channel, the only modification being that the output label of the edges in the error state diagram are calculated according to equation (\ref{eq_34}). Also notice that the zero cycles given in example \ref{exam_2} and example \ref{exam_3} remain the same in the asymmetric channel.

\begin{figure}
\centering
\hspace{-2pt}
\subfigure[$\epsilon=0.1$]{
\includegraphics[width=0.5\columnwidth]{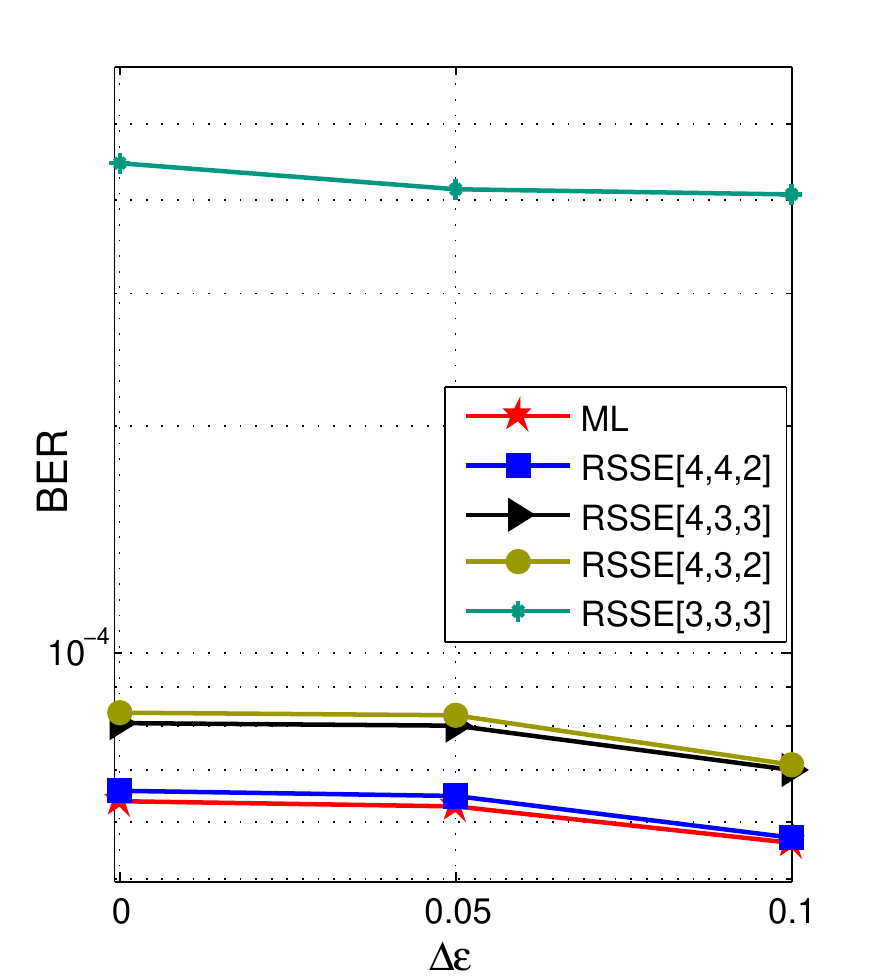}
\label{sub_asy_1}
}~
\hspace{-12pt}
\subfigure[$\epsilon=0.4$]{
\includegraphics[width=0.5\columnwidth]{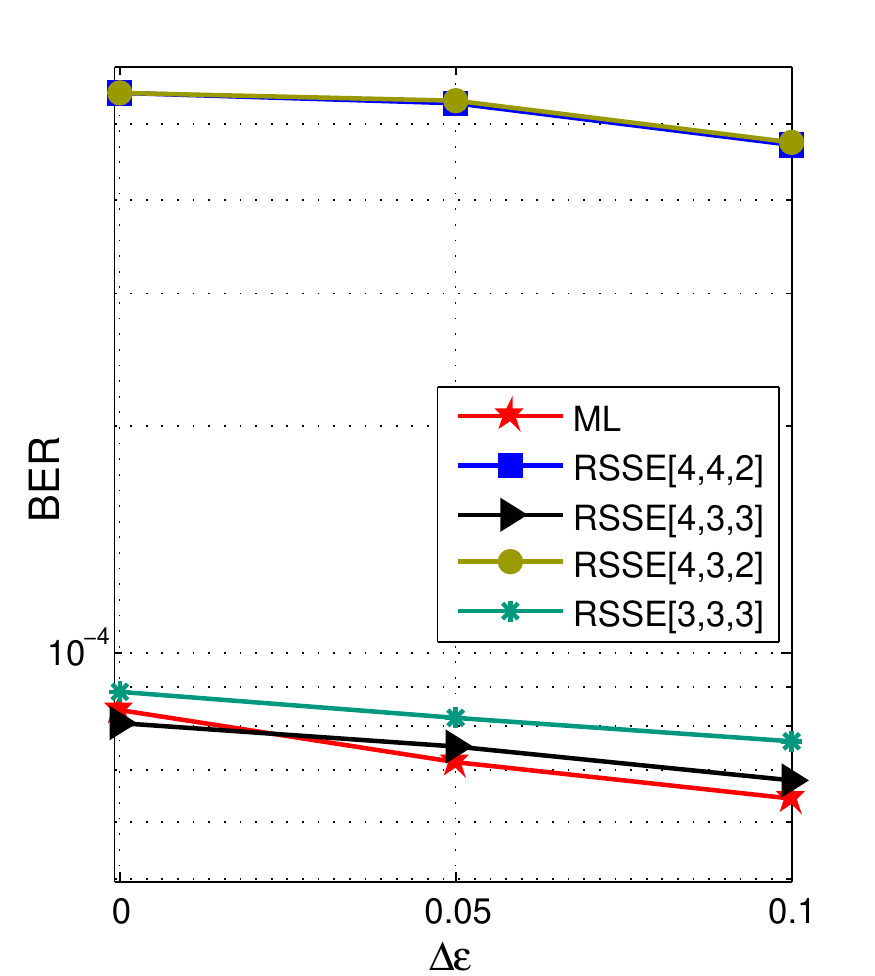}
\label{sub_asy_2}
}
\caption{Performance of RSSE on asymmetric 2H2T EPR4 channel, at different ITI levels and various offsets.}

\label{fig_asy}
\end{figure}
We search for $d^2_{\text{min}}(E^r)$ at ITI levels with two extreme values of $\epsilon$ and various offsets $\Delta\epsilon$ on the EPR4 channel. The results are listed in Table \ref{tab_asy1} and Table \ref{tab_asy2}.  In each case, $\Delta \epsilon$ could take values from $\{0, 0.05, 0.1\}$. For comparison, we also give the minimum distance parameter of the ML detector, denoted as $d^2_{\text{min}}$, in each corresponding scenario. We find that $d^2_{\text{min}}(E^r)$ does not change much from the symmetric case ($\Delta\epsilon=0$). In addition, some trellis configurations tend to have increased $d^2_{\text{min}}(E^r)$ under severe asymmetry, while some do not. We see that the performance of a configuration is closely related to the distance parameters of the length-1 error events, which provides an approach to design the set partition tree for other MHMT channel models. The conclusion is that the proposed RSSE algorithm is applicable to the asymmetric channel.
An example of BER vs. $\Delta\epsilon$ simulation result is shown in Figs. \ref{fig_asy}.

\section{Conclusion}
\label{sec_conclusion}
In this paper we deal with the problem of the high computational complexity of MHMT detection. The proposed RSSE algorithm with a modified set partition tree significantly reduces the number of trellis states while retaining near optimal performance. Performance was evaluated on both symmetric and asymmetric 2H2T channel models.  Error event analysis was used to explain the performance variation for different trellises under different ITI conditions. If a suitable set partition tree is designed, the RSSE algorithm has the potential to be applied to more general MHMT channels. This is a direction for future research.

\section*{Acknowledgment}
This work was supported in part by the National Science Foundation under Grant CCF-1405119, and the Center for Memory and Recording Research (formerly, Center for Magnetic Recording Research) at UC San Diego.

\bibliographystyle{IEEEtran}

\bibliography{bibliography_multitrack_phs}

\end{document}